\documentclass[Journal,twocolumn]{IEEEtran}
\IEEEoverridecommandlockouts
\usepackage{color}
\usepackage{graphicx}
\usepackage{amsmath}
\usepackage{amssymb}
\usepackage{algorithm}
\usepackage{algorithmic}
\usepackage{amsmath}
\usepackage{multirow}
\usepackage{booktabs}
\usepackage{array}
\usepackage{amsthm}
\usepackage{stfloats}
\usepackage{caption}
\usepackage{subfigure}
\usepackage{bm}
\usepackage{booktabs}
\usepackage{setspace}
\usepackage{url}
\usepackage{arydshln}
\usepackage{flushend}
\usepackage{comment}

\newcommand{\be}{\begin{equation}}
\newcommand{\ee}{\end{equation}}

\renewcommand{\thesubsubsection}{\thesubsection.\arabic{subsubsection}}

\renewcommand{\baselinestretch}{1}
\allowdisplaybreaks[4]

\title{Extended Target Sensing in MIMO-OFDM ISAC Systems: Modeling, Optimization and Estimation

\thanks{Part of this paper has been presented in IEEE International Radar Conference, 2025 \cite{RLiu-Radar-2025}.}
\thanks{R. Liu and A. L. Swindlehurst are with the Nhu Department of Electrical Engineering and Computer Science, University of California Irvine, CA 92697, USA (e-mail: rangl2@uci.edu; swindle@uci.edu).}
\thanks{M. Li is with the School of Information and Communication Engineering, Dalian University of Technology, Dalian 116024, China (e-mail: mli@dlut.edu.cn).}
}

\author{Rang Liu,~\IEEEmembership{Member,~IEEE,}
        Ming Li,~\IEEEmembership{Senior Member,~IEEE,}
        and A. Lee Swindlehurst,~\IEEEmembership{Life Fellow,~IEEE}}

\begin{document}

\maketitle
\pagestyle{empty}
\thispagestyle{empty}

\begin{abstract}
This paper develops a comprehensive target modeling, beamforming optimization, and parameter estimation framework for extended-target sensing in wideband MIMO-OFDM integrated sensing and communication systems. We propose a parametric scattering model (PSM) that decouples target geometry from electromagnetic scattering characteristics, requiring only six nonlinear geometric parameters and linear radar cross-section (RCS) coefficients. Based on this compact structure, we derive a hybrid Bayesian Cram\'{e}r-Rao bound (CRB) for joint estimation of azimuth, elevation, and range-related parameters. To handle inherent range ambiguities due to OFDM signaling, we analyze the range ambiguity function and introduce range sidelobe suppression constraints around the true range. Based on these constraints, we formulate an ambiguity-aware transmit beamforming design that minimizes a weighted geometric CRB subject to per-user signal-to-interference-plus-noise ratio (SINR) requirements and a total power budget. As benchmarks, we extend two other common models to the same wideband MIMO-OFDM scenario. We also derive maximum a posteriori estimators and a computational complexity analysis for all three models. Simulation results demonstrate that the proposed PSM-based approach achieves improved target localization with significantly reduced runtime for beamforming optimization and parameter estimation, while consistently satisfying communication SINR requirements.
\end{abstract}


\section{Introduction}

Integrated sensing and communication (ISAC) is a promising technology for sixth-generation (6G) wireless networks, enabling simultaneous high-throughput data transmission and accurate environmental perception \cite{FLiu-JSAC-2022}-\cite{Liu-WCM-2023}. Among various physical-layer techniques, multiple-input multiple-output orthogonal frequency division multiplexing (MIMO-OFDM)-based ISAC has attracted substantial attention due to its compatibility with current wireless standards and ability to exploit spatial and frequency degrees of freedom (DoFs) \cite{Sturm_Proc_2011}-\cite{QDai-arxiv-2025}. A fundamental challenge within this framework is transmit beamforming optimization, which aims to achieve an effective trade-off between sensing accuracy and communication quality-of-service (QoS). Existing research has considered various sensing performance metrics, including beampattern similarity \cite{XLiu-TSP-2020}, detection signal-to-interference-plus-noise ratio (SINR) \cite{RLiu-JSTSP-2022}-\cite{ZXiao-Tcom-2025}, and estimation-theoretic measures such as the Cram\'{e}r-Rao bound (CRB) \cite{Rang TWC 2024}, typically under communication constraints such as SINR or sum-rate requirements.

However, conventional ISAC beamforming designs often assume targets to be point-like scatterers characterized by a single dominant reflector. While this simplification is reasonable for traditional radar scenarios with relatively small long-range targets, it fails in practical applications like automotive radar and infrastructure monitoring, where the sensed objects extend across multiple resolution cells in angle and range. In systems employing large-scale millimeter-wave arrays with fine angular resolution, target echoes naturally span several azimuth-elevation-range bins. Ignoring the target's spatial extent can lead to biased localization and inefficient transmit beam patterns that poorly match the actual physical dimensions of extended targets. Therefore, accurate modeling of large targets is important for MIMO-OFDM ISAC systems.

The radar literature provides various extended-target models, often tailored towards receiver-side imaging or shape inference, including electromagnetic (EM)-based descriptions \cite{BSambon-TRS-2025} and detailed shape parameterizations \cite{ADaniyan-TSP-2018,Garcia-TSP-2022}. Although these high-fidelity models offer precise target characterization, they typically involve large numbers of parameters and computationally intensive algorithms, making them impractical for transmit beamforming optimization in ISAC scenarios. Recently, several ISAC studies have considered simplified geometric shape models such as contour-based reflections or truncated Fourier-series surfaces \cite{YWang-TWC-2024, YWang-arxiv-2025}, yet such models lead to nonconvex estimation problems whose solutions result in significant performance gaps compared to theoretical bounds.
The NR-V2X beam-management approach in \cite{SZhou-arxiv-2026} maps angle-domain CRBs to predictive error ellipses around resolvable dominant scatterers. Based on this, access and tracking beams are designed to cover the induced uncertainty set via a minimum-enclosing-ellipse with half-power beamwidth control. This formulation reduces alignment overhead, but it primarily captures angular uncertainty and typically relies on scatterer resolvability.

Motivated by the need for tractable modeling and optimization, most existing transmit-oriented ISAC designs adopt simplified ``optimization-friendly'' models, primarily the unstructured channel model (UCM) and the discrete scattering model (DSM). The UCM represents the extended-target response as an arbitrary unstructured matrix, allowing direct optimization of matrix entries \cite{FLiu-TSP-2021}-\cite{XSong-TWC-2024}. While this abstraction significantly simplifies beamforming optimization, it neglects geometric constraints embedded in array manifolds, resulting in diffuse beampatterns and introducing many unnecessary variables. In addition, UCM-based approaches primarily focus on CRB-oriented transmit design and only infer target geometric parameters from an estimated response matrix, without providing matched end-to-end receiver algorithms. On the other hand, the DSM represents an extended target by discretizing it into individual scatterers distributed across angle and range \cite{ZDu-TWC-2023}-\cite{PMaity-OJCS-2025}. While maintaining clear physical interpretability, DSM-based approaches involve high-dimensional nonlinear optimization and parameter estimation at the scatterer-level, making them computationally demanding, especially in wideband MIMO-OFDM scenarios. Thus, an intermediate target model is needed, one that simultaneously maintains the geometric interpretability of DSM and the tractable dimensionality of UCM, while supporting matched receiver algorithms suitable for end-to-end performance evaluation.

Beyond accurate modeling, realistic extended‑target sensing in wideband MIMO‑OFDM ISAC systems poses additional challenges in both transmit beamforming optimization and receiver‑side parameter estimation. Most existing CRB‑oriented beamforming designs are developed for narrowband or single‑carrier systems \cite{SKarbasi-TSP-2015}-\cite{ YYao-TITS-2022}, and therefore do not fully exploit the structure of wideband multi‑carrier signaling. Consequently, these methods often cannot adequately generalize to practical three-dimensional (3D) extended targets. Furthermore, the CRB inherently characterizes only local estimation accuracy around the true parameters. Due to the periodic phase structure across OFDM subcarriers, the range ambiguity function exhibits multiple peaks and pronounced sidelobes \cite{MFKeskin-TSP-2021,Peishi-TWC-2025}, limiting the effectiveness of beamforming strategies designed solely based on local CRB minimization. Equally important is receiver‑side parameter estimation: without matched estimation algorithms of manageable complexity, the theoretical advantages suggested by CRB‑based transmit designs are difficult to realize in practice. Existing UCM‑ and DSM‑based approaches typically rely on high‑dimensional intermediate estimation, such as reconstructing unstructured channel responses or estimating scatterer‑level parameters prior to geometric parameter extraction. This two‑stage processing incurs substantial computational cost and is prone to information loss and error propagation.

To address these issues, we propose a unified modeling, optimization, and estimation framework for extended‑target sensing in wideband MIMO‑OFDM ISAC systems. The core of the proposed approach is a parametric scattering model (PSM) that provides a compact and optimization‑friendly representation of extended targets. Based on this model, we develop an ambiguity‑aware CRB‑oriented transmit beamforming design and matched receiver‑side estimation algorithms. The main contributions of the paper are summarized below.
\begin{itemize}
\item \textbf{Modeling:} We propose a compact PSM for extended targets that separates a small set of geometric parameters from the linear radar cross-section (RCS) coefficients. This representation substantially reduces the nonlinear parameter dimension compared to DSM, while retaining geometric interpretability that is absent in UCM. Under a unified wideband MIMO-OFDM setting, we further relate the CRBs of UCM- and DSM-based formulations to the geometric parameter space through Jacobian mappings, enabling consistent and fair performance comparisons across different modeling paradigms.

\item \textbf{Optimization:} To address the inherent range ambiguity arising from wideband OFDM signaling, we analyze the associated ambiguity function and introduce practical range-sidelobe suppression constraints around the true target range. With these constraints, we formulate an ambiguity-aware transmit beamforming optimization problem that minimizes a hybrid Bayesian CRB for joint azimuth, elevation, and range parameter estimation, subject to per-user SINR and total transmit power constraints. The resulting optimization framework effectively mitigates ambiguity issues in multiuser wideband MIMO-OFDM ISAC systems.

\item \textbf{Estimation:} We develop matched parameter estimation algorithms not only for the proposed PSM but also for the UCM and DSM frameworks, to fully describe the receiver-side processing commonly overlooked in the existing CRB-based ISAC literature. Specifically, we investigate two-stage estimation procedures and utilize the extended invariance principle (EXIP) \cite{EXIP1}-\cite{EXIP3} to optimally extract geometric parameters from high-dimensional intermediate estimates. 
The proposed PSM estimator maintains a low-dimensional nonlinear parameterization, significantly reducing computational complexity. Simulation results further demonstrate that our proposed approach closely approaches the theoretical CRB while substantially reducing computational complexity compared to UCM- and DSM-based methods.
\end{itemize}

\emph{Notation:} 
Scalars, vectors, and matrices are represented by italic letters $x$, bold lowercase letters $\mathbf{x}$, and bold uppercase letters $\mathbf{X}$. The sets of real and complex matrices of size $m\times n$ are denoted by $\mathbb{R}^{m\times n}$ and $\mathbb{C}^{m\times n}$, respectively. The set of $n\times n$ Hermitian positive semidefinite matrices is denoted by $\mathbb{S}_+^n$. The $T\times 1$ all-ones vector and $n\times n$ identity matrix are denoted by $\mathbf{1}_T$ and $\mathbf{I}_n$. Operators $(\cdot)^T$, $(\cdot)^H$, and $(\cdot)^*$ denote transpose, Hermitian transpose, and complex conjugation. $\text{Tr}\{\cdot\}$, $\text{vec}\{\cdot\}$, $\text{rank}(\cdot)$, $\text{diag}\{\cdot\}$, and $\text{blkdiag}\{\cdot\}$ denote trace, vectorization, rank, diagonal, and block-diagonal matrices, respectively. $\Re\{\cdot\}$ and $\Im\{\cdot\}$ extract real and imaginary parts, and $|\cdot|$ is the absolute value. $\|\cdot\|_2$ and $\|\cdot\|_F$ denote the Euclidean and Frobenius norms. Expectation is represented by $\mathbb{E}\{\cdot\}$, and $\mathcal{CN}(\bm{\mu},\mathbf{C})$ and $\mathcal{N}(\bm{\mu},\mathbf{C})$ represent complex and real Gaussian distributions. The notation $\mathbf{A}\succeq\mathbf{0}$ means that $\mathbf{A}$ is positive semidefinite, $\otimes$ denotes the Kronecker product, $\lceil \cdot \rceil$ is the ceiling function, and $\jmath$ is the imaginary unit ($\jmath^2=-1$).

\section{System and Target Scattering Model}

We consider a monostatic ISAC base station (BS) comprising separate transmit and receive uniform rectangular arrays (URAs) with $N_\text{t}=N_\text{tx} \times N_\text{ty}$ and $N_\text{r}=N_\text{rx} \times N_\text{ry}$ antennas, respectively. The BS simultaneously serves $K$ single-antenna users and senses an extended target.

\subsection{Transmit Signal Model and Communication Performance}

The BS employs OFDM with $N$ subcarriers indexed by $n\in\{0,1,\dots,N-1\}$ and transmits $L$ OFDM symbols indexed by $l\in\{0,1,\dots,L-1\}$ within each coherent processing interval (CPI). The symbol duration is defined as $T_\text{sym}=1/\Delta_f + T_\text{CP}$, where $\Delta_f$ denotes the subcarrier spacing and $T_\text{CP}$ represents the cyclic prefix duration. The frequency-domain transmit signal on the $n$-th subcarrier during the $l$-th OFDM symbol, denoted by $\mathbf{x}_n[l]\in\mathbb{C}^{N_\text{t}}$, is generated through  linear precoding applied to a combination of communication symbols $\mathbf{s}_{\text{c},n}[l]\in\mathbb{C}^K$ and sensing-specific signals $\mathbf{s}_{\text{s},n}[l]\in\mathbb{C}^{N_\text{t}}$. Thus, we have: 
\begin{align}
\mathbf{x}_n[l] = \mathbf{W}_{\text{c},n}\mathbf{s}_{\text{c},n}[l] + \mathbf{W}_{\text{s},n}\mathbf{s}_{\text{s},n}[l] = \mathbf{W}_n\mathbf{s}_n[l],
\end{align}
where $\mathbf{W}_{\text{c},n}\in\mathbb{C}^{N_\text{t}\times K}$ and $\mathbf{W}_{\text{s},n}\in\mathbb{C}^{N_\text{t}\times N_\text{t}}$ represent the communication and sensing beamforming matrices, respectively. The combined beamforming matrix is $\mathbf{W}_n=[\mathbf{W}_{\text{c},n},~\mathbf{W}_{\text{s},n}]\in\mathbb{C}^{N_\text{t}\times (N_\text{t}+K)}$, and the stacked symbol vector is defined as $\mathbf{s}_n[l]=[\mathbf{s}_{\text{c},n}[l]^T,\mathbf{s}_{\text{s},n}[l]^T]^T\in\mathbb{C}^{N_\text{t}+K}$. 
The $N_\text{t}$ sensing-specific streams enable flexible transmit covariance optimization for enhanced sensing performance.
The symbols are assumed independent with unit power: $\mathbb{E}\{\mathbf{s}_n[l]\mathbf{s}^H_n[l]\}=\mathbf{I}_{N_\text{t}+K}$. 

The signal at the $k$-th user on the $n$-th subcarrier is
\begin{align}
y_{n,k}[l] = \mathbf{h}_{n,k}^H\mathbf{W}_n\mathbf{s}_n[l] + z_{n,k}[l],
\end{align}
where $\mathbf{h}_{n,k}\in\mathbb{C}^{N_\text{t}}$ denotes the frequency-domain communication channel from the BS to the $k$-th user, and $z_{n,k}[l]\sim\mathcal{CN}(0,\sigma_\text{c}^2)$ is additive white Gaussian noise (AWGN). Consequently, the SINR for the $k$-th user on subcarrier $n$ is  
\begin{align}
\text{SINR}_{n,k} = \frac{|\mathbf{h}_{n,k}^H\mathbf{w}_{n,k}|^2}
{\sum_{j\neq k}^{N_\text{t}+K}|\mathbf{h}_{n,k}^H\mathbf{w}_{n,j}|^2+\sigma_\text{c}^2},
\end{align}
where $\mathbf{w}_{n,j}$ denotes the $j$-th column of $\mathbf{W}_n$. Note that the dedicated sensing streams are regarded as additional interference and thus jointly optimized with the communication signals under SINR constraints.

\subsection{Unstructured Channel Model (UCM) for Extended Target}

We first consider the most general representation of the extended-target echo channel, known as the UCM. Under the UCM framework, the frequency-domain echo at the BS on the $n$-th subcarrier during the $l$-th OFDM symbol is modeled as
\begin{align}
\mathbf{y}_n[l] = \mathbf{G}_n\mathbf{x}_n[l]+ \mathbf{z}_n[l],
\end{align}
where $\mathbf{G}_n\in\mathbb{C}^{N_\text{r}\times N_\text{t}}$ denotes the frequency-dependent target response matrix, and $\mathbf{z}_n[l]\in\mathbb{C}^{N_\text{r}}\sim\mathcal{CN}(\mathbf{0},\sigma_\text{s}^2\mathbf{I}_{N_\text{r}})$ represents the receiver noise.

The UCM treats each subcarrier-dependent channel matrix $\mathbf{G}_n$ as an arbitrary and unconstrained matrix. Consequently, this model requires direct estimation of a large number of parameters, specifically $NN_\text{r}N_\text{t}$ complex channel coefficients across all subcarriers, significantly increasing computational complexity and limiting interpretability. The absence of geometric constraints in UCM generally results in diffuse beampatterns and reduced estimation efficiency, motivating more structured modeling approaches.

\subsection{Discrete Scattering Model (DSM) for Extended Target}

To incorporate the target geometry and reduce dimensionality, the DSM represents an extended target as a three-dimensional collection of discrete scatterers. The target region is discretized into a spatial grid composed of $T = T_\theta\times T_\phi \times T_d$ scatterers, where $T_\theta$, $T_\phi$, and $T_d$ denote the numbers of discretization points along the azimuth, elevation, and range dimensions, respectively. Each scatterer is indexed by the triplet $(p,q,r)$ with $p\in\{1,2,\dots,T_\theta\}$, $q\in\{1,2,\dots,T_\phi\}$, and $r\in\{1,2,\dots,T_d\}$, which is then mapped to a single linear index $t=(p-1)T_\phi T_d+(q-1)T_d+r$. Thus, each scatterer has an associated azimuth angle $\theta_t$, elevation angle $\phi_t$, range $d_t$, and a complex coefficient $\alpha_t\in\mathbb{C}$ that encapsulates RCS characteristics and distance-dependent attenuation.

Under this framework, the target response matrix $\mathbf{G}_n$ can be represented as the sum of contributions from all scatterers: 
\begin{align}\label{eq:Gn}
    \mathbf{G}_n = \sum_{t=1}^T\alpha_tf_{t,n}\mathbf{b}_{t,n}\mathbf{a}_{t,n}^H,
\end{align}
where $f_{t,n} =e^{-\jmath2\pi n\Delta_f\tau_t}$ with round-trip delay $\tau_t=2d_t/c$, and $\mathbf{a}_{t,n}\in\mathbb{C}^{N_\text{t}}$ and  $\mathbf{b}_{t,n}\in\mathbb{C}^{N_\text{r}}$ denote the transmit and receive steering vectors. The target is assumed to lie in the far field of the BS and remain approximately static within the CPI. For the URA configuration, the steering vectors admit a Kronecker product decomposition: 
\begin{align} \label{eq:abvec}
\mathbf{a}_{t,n} = \mathbf{a}_{\text{x},n}(\theta_t,\phi_t)\otimes \mathbf{a}_{\text{y},n}(\theta_t,\phi_t), 
\end{align} 
where the horizontal and vertical components are 
\begin{subequations}\label{eq:axay}
\begin{align} 
\mathbf{a}_{\text{x},n}(\theta,\phi)&\!=\![1,e^{-\jmath\pi\chi_n\sin\phi\cos\theta},\dots,e^{-\jmath(N_\text{tx}\!-\!1)\pi\chi_n\sin\phi\cos\theta}]^T,\\
\mathbf{a}_{\text{y},n}(\theta,\phi)&\!=\![1,e^{-\jmath\pi\chi_n\sin\phi\sin\theta},\dots,e^{-\jmath(N_\text{ty}\!-\!1)\pi\chi_n\sin\phi\sin\theta}]^T.
\end{align}\end{subequations}
A similar structure applies to $\mathbf{b}_{t,n}$. The scalar $\chi_n = 1+n\Delta_f/f_\text{c}$ accounts for subcarrier-dependent wavelength variation, which is significant in wideband OFDM sensing. When $\Delta_f/f_\text{c}\ll 1$, we have $\chi_n\approx 1$ and the steering vectors reduce to their narrowband forms.

Although the DSM in \eqref{eq:Gn} provides a physical interpretation and explicitly leverages the known array geometry, it still faces computational challenges due to its high dimensionality, particularly for finely discretized target regions, with $3T$ geometric parameters (azimuth, elevation, range for each scatterer) and $2T$ real-valued RCS parameters. This limitation motivates a more compact yet physically interpretable model.

\subsection{Proposed Parametric Scattering Model (PSM)} 

To effectively overcome the computational burden caused by high dimensionality in DSM and the lack of structural constraints in UCM, we propose a parametric scattering model (PSM) that elegantly combines intrinsic geometric constraints with flexible modeling of local scattering variations.

The proposed PSM characterizes an extended target using only six geometric parameters: the center coordinates $(\theta_0,~\phi_0,~d_0)$ and the extent of the target in the angular-range domains $(\Delta_\theta,~\Delta_\phi,~\Delta_d)$. Based on the discretization approach introduced in Section II-C, each scatterer with index $(p,q,r)$ is placed within the angular-range domain according to: 
\begin{subequations}\label{eq:geom pqr}
\begin{align}
\theta_p &= \theta_0 + \Delta_\theta u_p,\quad u_p \triangleq 
\frac{p-1}{T_\theta-1}-\frac12, & T_\theta > 1, \\
\phi_q &= \phi_0 + \Delta_\phi v_q,\quad v_q \triangleq 
\frac{q-1}{T_\phi-1}-\frac12, & T_\phi > 1,\\
d_r & = d_0 + \Delta_d w_r,\quad w_r \triangleq 
\frac{r-1}{T_d-1}-\frac12, & T_d > 1.
\end{align}\end{subequations}
This ensures uniform and systematic sampling across the entire defined angular-range extent of the target. Each scatterer's position $(\theta_t, \phi_t, d_t)$ is then uniquely identified through the single-index mapping $t=(p-1)T_\phi T_d+(q-1)T_d+r$, and all scatterer positions are collectively determined by the compact geometric parameter vector:
\begin{align} \label{eq:xi geom}
\bm{\xi}_\text{geom}=[\theta_0,~\Delta_\theta,~\phi_0,~\Delta_\phi,~d_0,~\Delta_d]^T \in \mathbb{R}^6. 
\end{align} 
While the target scatterers are geometrically constrained by $\bm{\xi}_\text{geom}$, their associated complex reflection coefficients $\bm{\alpha} \triangleq [\alpha_1,\alpha_2,\dots,\alpha_T]^T$ are not assumed to be a function of $\bm{\xi}_\text{geom}$. To capture inherent scattering randomness, we adopt a Gaussian prior for the complex coefficients:  $\bm{\alpha}\sim\mathcal{CN}(\mathbf{0},\sigma_\alpha^2\mathbf{I}_T)$. Consequently, the complete PSM parameter vector is: 
\begin{align}\label{eq:xi} 
\bm{\xi}=\left[\bm{\xi}_\text{geom}^T,~\Re\{\bm{\alpha}^T\},~\Im\{\bm{\alpha}^T\}\right]^T\in\mathbb{R}^{2T+6}.
\end{align}

Under this parameterization, the target response matrix $\mathbf{G}(\bm{\xi})$ retains the DSM superposition form \eqref{eq:Gn}, with scatterer coordinates defined by $\bm{\xi}_\text{geom}$ through \eqref{eq:geom pqr}. Thus, the geometric parameters influence the steering vectors nonlinearly, whereas the RCS coefficients enter linearly. Given $\mathbf{G}(\bm{\xi})$, the estimate of $\bm{\alpha}$ can be efficiently computed through weighted least squares, effectively decoupling parameter estimation into a low-dimensional nonlinear optimization followed by a straightforward linear update.
Compared to DSM and UCM, the proposed PSM significantly reduces the nonlinear optimization dimension from $3T$ to six geometric parameters, independent of discretization granularity. This substantial dimensionality reduction markedly enhances computational tractability while maintaining robust modeling flexibility.

Effective use of the proposed PSM requires a principled discretization strategy balancing estimation accuracy and computational complexity. The discretization resolutions in azimuth, elevation, and range,  denoted by $\delta_\theta$, $\delta_\phi$, $\delta_d$, are determined by the array geometry and signal bandwidth \cite{HTrees-2002}:
\begin{align}\label{eq:resolution}
\delta_\theta \approx \frac{2}{N_\text{tx}N_\text{rx}\sin\phi_0}, ~~\delta_\phi \approx  \frac{2}{N_\text{ty}N_\text{ry}\cos\phi_0 },  ~~ \delta_d =  \frac{c}{2N\Delta_f}.
\end{align}
Given these resolutions and the target size $\Delta_\theta$, $\Delta_\phi$, $\Delta_d$, the discretization points $(T_\theta, T_\phi, T_d)$ are chosen as 
\begin{align} \label{eq:T_x}
T_x = \max\left\{1, \left\lceil\Delta_x/\delta_x\right\rceil+1\right\},\quad x \in \{\theta,~\phi,~d\}.
\end{align}

To summarize, compared to the UCM, which treats the subcarrier response matrices as entirely unstructured, the DSM reduces complexity by modeling scatterer-level geometry but still suffers from high-dimensional parameter estimation. The proposed PSM further simplifies the model by separating geometry and scattering effects into just six nonlinear geometric parameters and $T$ linear RCS coefficients. This significantly reduces complexity and preserves interpretability, providing a manageable balance between accuracy and computational tractability. In subsequent sections, we derive CRBs and design CRB-oriented beamformers for these modeling frameworks.

\section{PSM-Based Beamforming Optimization for Extended Target Sensing}
Having established the PSM framework, we now develop a CRB‑oriented transmit beamforming design. The compact geometric parameterization of the PSM enables an efficient optimization procedure while preserving sufficient flexibility to represent extended targets.

\subsection{CRB Derivation for Geometric Parameters}

To characterize the fundamental estimation limits for extended targets in the considered MIMO‑OFDM ISAC system, we derive a CRB based on the proposed parametric scattering model. We begin with the wideband observation model across all $L$ OFDM symbols and $N$ subcarriers. At subcarrier $n$, the received signal matrix $\mathbf{Y}_n \triangleq \big[\mathbf{y}_n[0],\mathbf{y}_n[1],\dots,\mathbf{y}_n[L-1]\big]\in\mathbb C^{N_\text{r}\times L}$ is expressed as 
\begin{align}
\mathbf{Y}_n = \mathbf{G}_n \mathbf{X}_n + \mathbf{Z}_n,
\end{align}
where $\mathbf{X}_n \triangleq  [\mathbf{x}_n[0],\mathbf{x}_n[1],\dots,\mathbf{x}_n[L-1]]\in\mathbb C^{N_\text{t}\times L}$ denotes the transmitted waveform matrix constructed as $\mathbf{X}_n = \mathbf{W}_n\mathbf{S}_n$ with the symbol matrix $\mathbf S_n = [\mathbf{s}_n[0],\dots,\mathbf{s}_n[L-1]]
  \in\mathbb C^{(N_\text{t}+K)\times L}$, and the noise matrix $\mathbf{Z}_n = [\mathbf{z}_n[0],\mathbf{z}_n[1],\dots,\mathbf{z}_n[L-1]]\in\mathbb C^{N_\text{r}\times L}$.
Stacking all subcarrier signals yields the full wideband observation model
\begin{align}\label{eq:observe Y}
\mathbf{Y} = \mathbf{GX} + \mathbf{Z},
\end{align} 
where $\mathbf{Y} \triangleq \left[\mathbf{Y}_0,\mathbf{Y}_1,\dots,\mathbf{Y}_{N-1}\right]\in\mathbb C^{N_\text{r}\times NL}$, $\mathbf{G} \triangleq[\mathbf{G}_0,\mathbf{G}_1,\dots,\mathbf{G}_{N-1}]\in\mathbb{C}^{N_\text{r}\times NN_\text{t}} 
$ represents the extended target response parameterized by $\bm{\xi}$, the waveform matrix has the block diagonal structure: 
$\mathbf{X} \triangleq \text{blkdiag}\{\mathbf{X}_0,\mathbf{X}_1,\dots,\mathbf{X}_{N-1}\}$, and $\mathbf{Z} \triangleq [\mathbf{Z}_0,\mathbf{Z}_1,\dots,\mathbf{Z}_{N-1}]$ aggregates the noise components.
Vectorizing \eqref{eq:observe Y} gives
\begin{align}\label{eq:CRB_vec_sig} 
\mathbf{y} \triangleq \text{vec}\{\mathbf{Y}\} = \text{vec}\{\mathbf{G}(\bm{\xi})\mathbf{X}\}+\mathbf{z}
= \bm{\eta}(\bm{\xi}) + \mathbf{z}, 
\end{align} 
with $\mathbf{z} = \text{vec}\{\mathbf{Z}\}\sim\mathcal{CN}(\mathbf{0},\sigma_\text{s}^2\mathbf{I}_{LNN_\text{r}})$.
Note that the model is linear in $\bm{\alpha}$, i.e., $\mathbf{y} = \mathbf{D}(\bm{\xi}_\text{geom})\bm{\alpha} + \mathbf{z}$, where the dictionary matrix $\mathbf{D}(\bm{\xi}_\text{geom})$ is defined as 
\begin{align}\label{eq:Dxi}
\mathbf{D}(\bm{\xi}_\text{geom})\triangleq\big[\text{vec}\{\mathbf{V}_1(\bm{\xi}_\text{geom})\mathbf{X}\},\dots,\text{vec}\{\mathbf{V}_T(\bm{\xi}_\text{geom})\mathbf{X}\}\big],    
\end{align}
where $\mathbf{V}_t = [\mathbf{V}_{t,0},\dots,\mathbf{V}_{t,N-1}]$ with $\mathbf{V}_{t,n} = f_{t,n}\mathbf{b}_{t,n}\mathbf{a}_{t,n}^H$.

We treat the geometric parameters $\bm{\xi}_\text{geom}$ defined in~\eqref{eq:xi geom} as deterministic unknowns since they typically vary slowly within a CPI. In contrast, the complex RCS coefficients $\bm{\alpha}$ are assumed to exhibit local fluctuations and thus are modeled as independent Gaussian random variables $\bm{\alpha}\sim\mathcal{CN}(\mathbf{0},\sigma_\alpha^2\mathbf{I}_T)$. This motivates a hybrid Bayesian CRB formulation, leading to the following FIM decomposition:
\begin{equation}\label{eq:hybrid FIM}
\mathbf{F} = \mathbb{E}_{\bm{\alpha}}\!\left\{\mathbf{F}_{\text{data}}(\bm{\xi}_{\text{geom}}, \bm{\alpha})\right\}
    + \mathbf{F}_{\text{prior}},   
\end{equation}
where the prior term $\mathbf{F}_{\text{prior}}$ impacts only the block associated with the RCS parameters.
The hybrid FIM can be partitioned into blocks corresponding to geometric and RCS parameters: 
\begin{align}\label{eq:FIM_block} 
\mathbf{F}=\begin{bmatrix} \mathbf{F}_\text{gg} & \mathbf{F}_{\text{g}\alpha}\vspace{0.15 cm} \\ \mathbf{F}_{\text{g}\alpha}^T & \mathbf{F}_{\alpha\alpha} \end{bmatrix} , 
\end{align} 
where $\mathbf{F}_\text{gg}\in\mathbb{R}^{6\times 6}$, $\mathbf{F}_{\alpha\alpha}\in\mathbb{R}^{2T\times 2T}$, and $\mathbf{F}_{\text{g}\alpha}\in\mathbb{R}^{6\times 2T}$ represent geometric, RCS, and cross-information blocks, respectively. The $\mathbf{F}_{\text{g}\alpha}$ term vanishes upon expectation with respect to $\bm{\alpha}$, since $\bm{\eta}(\bm{\xi})$ is linear in $\bm{\alpha}$ but nonlinear in $\bm{\xi}_\text{geom}$.
Therefore, the entries of $\mathbf{F}_{\text{g}\alpha}$ involve expectations of the form $\mathbb{E}_{\bm{\alpha}}\{\bm{\alpha}\}=\mathbf{0}$, and no additional cross-terms appear since the geometric and RCS parameters are independent. Thus, the geometric parameters decouple from $\bm{\alpha}$, simplifying the CRB for the geometric parameters as
\begin{align}\label{eq:CRB_geom} 
\mathbf{C}_\text{geom}= \mathbf{F}_\text{gg}^{-1}. 
\end{align}

From \eqref{eq:CRB_vec_sig}, each entry of $\mathbf{F}_{\text{gg}}$ is computed as 
\begin{subequations}
\begin{align}\label{eq:Fij}
\mathbf{F}_{\text{gg}}(i,j) &= \frac{2}{\sigma_\text{s}^2}\Re\Big\{\frac{\partial^H\bm{\eta}}{\partial\xi_i}\frac{\partial\bm{\eta}}{\partial\xi_j}\Big\}\\
\label{eq:Fij define}
&= \sum_{n=0}^{N-1}\Re\left\{\text{Tr}\left\{\mathbf{R}_n\mathbf{C}_{i,j,n}\right\}\right\},
\end{align} \end{subequations}
where the transmit covariance matrices are defined as  
\begin{align}
\mathbf{R}_n = \frac{1}{L}\mathbb{E}\{\mathbf{X}_n\mathbf{X}_n^H\} = \mathbf{W}_n\mathbf{W}_n^H,
\end{align}
and the matrices $\mathbf{C}_{i,j,n}$ incorporate parameter-specific contributions at subcarrier $n$ (defined in detail in Appendix~A).

Finally, to construct a scalar metric suitable for beamforming optimization, we adopt a weighted trace criterion based on the geometric CRB:
\begin{align} \label{eq:WBCRB}
    f = \text{Tr}\{\bm{\Lambda}\mathbf{C}_{\text{geom}}\} = \text{Tr}\{\bm{\Lambda}\mathbf{F}_{\text{gg}}^{-1}\},  
\end{align}
where the diagonal weighting $\bm{\Lambda} \triangleq \text{diag}\{\lambda_1^2,\lambda_2^2,\dots,\lambda_6^2\}$ ensures appropriate normalization of the geometric parameters according to their inherent resolutions defined in \eqref{eq:resolution}. This weighting avoids emphasis on parameters with larger numerical values and enables a balanced consideration of angular and range estimation accuracy.

While the CRB quantifies local estimation accuracy, it does not account for the structure induced by the periodic range-dependent phase of OFDM waveforms. Specifically, the likelihood or matched-filter response in range typically exhibits significant secondary peaks spaced approximately by $c/(2\Delta_f)$, even when the local CRB is small. Moreover, CRB-oriented beamforming designs tend to concentrate transmit power onto sparse subsets of subcarriers to improve local sensitivity, inadvertently increasing effective subcarrier spacing and amplifying nearby ambiguity function sidelobes. Consequently, relying solely on minimizing the CRB may heighten the risk of confusion among closely spaced range hypotheses. To mitigate this effect, we incorporate ambiguity-function-based constraints that suppress sidelobes in the vicinity of the true range. These constraints are derived and integrated into the optimization framework in the following subsection.

\subsection{PSM Ambiguity Function and Range Sidelobe Constraints}

To address the range ambiguity described above, we introduce ambiguity-function-based constraints designed specifically for the proposed PSM. These constraints focus on selectively suppressing critical range sidelobes near the true target parameters. The ambiguity function quantifies the matched-filter response between signals corresponding to the true and hypothesized target parameters, defined as:
\begin{align}\label{eq:ambiguity_def}
\chi(\bm{\xi}_\text{geom}, \hat{\bm{\xi}}_\text{geom}) = \Big|\!\sum_{n=0}^{N-1}\text{Tr}\{\mathbf{R}_n \mathbf{G}_n^H(\bm{\xi}_\text{geom})\mathbf{G}_n(\hat{\bm{\xi}}_\text{geom})\}\Big|^2.
\end{align}
Under the assumption $\bm{\alpha} \sim \mathcal{CN}(\mathbf{0}, \sigma_\alpha^2 \mathbf{I}_T)$, the cross-correlation term $\mathbf{A}_n\triangleq \mathbf{G}_n^H(\hat{\bm{\xi}}_\text{geom})\mathbf{G}_n(\bm{\xi}_\text{geom})$ simplifies to  
\begin{subequations}\begin{align}
\mathbf{A}_n &= \sigma_\alpha^2 \sum_{t=1}^{T} \mathbf{V}_{t,n}^H(\bm{\xi}_\text{geom}) \mathbf{V}_{t,n}(\hat{\bm{\xi}}_\text{geom})    \\
&  = \sigma_\alpha^2 \sum_{t=1}^{T}f_{t,n}^* \hat{f}_{t,n} \mathbf{b}_{t,n}^H \hat{\mathbf{b}}_{t,n} \mathbf{a}_{t,n} \hat{\mathbf{a}}_{t,n}^H. 
\end{align}\end{subequations}


To illustrate the range ambiguity, we evaluate the ambiguity function along the range dimension $d_0$, fixing all other geometric parameters as $\hat{\bm{\xi}}_\text{geom} = [\theta_0, \Delta_\theta, \phi_0, \Delta_\phi, \hat{d}_0, \Delta_d]^T$. Under this simplification, the phase correlation term reduces to 
\begin{align}
f_{t,n}^* \hat{f}_{t,n} = \exp\{-\jmath 2\pi n\Delta_f \cdot 2(d_0 - \hat{d}_0)/c\}.
\end{align}
Hence, the simplified range ambiguity function becomes:
\begin{align}\label{eq:range_AF}
\chi_{d_0}(\hat{d}_0) = \sigma_\alpha^4 N_\text{r}^2 \Big|\sum_{n=0}^{N-1} e^{-\jmath\psi_n(\hat{d}_0)}\sum_{t=1}^{T} \mathbf{a}_{t,n}^H \mathbf{R}_n \mathbf{a}_{t,n}\Big|^2,
\end{align}
where $\psi_n(\hat{d}_0) = 2\pi n\Delta_f \cdot 2(d_0-\hat{d}_0)/c$.

To construct practical and interpretable constraints, we normalize the ambiguity function by its ideal mainlobe intensity evaluated at the true range $d_0$. Under maximum coherent integration and optimal beam alignment, this ideal mainlobe amplitude is approximately: $\chi_{d_0}(d_0) \approx \sigma_\alpha^4 N_\text{r}^2N_\text{t}^2P^2T^2$, where $P$ is the transmit power. Based on this normalization, the sidelobe suppression constraints at selected discrete points around the true range become: 
\begin{align}\label{eq:PSM sidelobe const}
\frac{\chi_{d_0}(\hat{d}_0)}{\chi_{d_0}(d_0)} \leq \epsilon_{d_0}, ~ \forall  \hat{d}_0 \in  \mathcal{D}_\text{side}^{d_0},
\end{align}
where the set $\mathcal{D}_\text{side}^{d_0}$ contains critical discrete points within a carefully chosen neighborhood around $d_0$. These points typically span several times the intrinsic range resolution interval $\delta_d$. The threshold $\epsilon_{d_0}>0$ controls the sidelobe suppression level. The detailed optimization formulation incorporating these constraints is presented in the following subsection.

\subsection{Formulation of PSM-Based Beamforming Design}

Building on the CRB analysis and ambiguity-function constraints established above, we now formulate the transmit beamforming optimization problem. Specifically, we design the beamforming matrices $\{\mathbf{W}_n\}_{n=0}^{N-1}$ to minimize the weighted CRB defined in \eqref{eq:WBCRB}, while enforcing the sidelobe suppression constraints in \eqref{eq:PSM sidelobe const} to ensure robust range estimation performance. This optimization is further subject to per-user SINR requirements $\Gamma_{n,k}$ for communication quality-of-service and a total transmit power budget $P$. Mathematically, the optimization problem is formulated as follows:
\begin{subequations}\label{eq:original problem} \begin{align} \min_{\{\mathbf{W}_n\}_{n=0}^{N-1}} & \quad \text{Tr}\left\{\bm{\Lambda}\mathbf{F}_\text{gg}^{-1}\right\} \label{eq:CRB_obj}\\
\label{eq:ambiguity_constraint}
\text{s.t.} &\quad \chi_{d_0}(\hat{d}_0)/\chi_{d_0}(d_0) \leq \epsilon_{d_0}, \quad \forall  \hat{d}_0 \in  \mathcal{D}_\text{side}^{d_0}, \\
&\quad \frac{|\mathbf{h}_{n,k}^H\mathbf{w}_{n,k}|^2}{\sum_{j\neq k}^{K+N_{\text{t}}}|\mathbf{h}_{n,k}^H\mathbf{w}_{n,j}|^2+\sigma_{\text{c}}^2}\geq\Gamma_{n,k},~\forall n, k,\label{eq:SINR_constraint}\\
&\quad \sum_{n=0}^{N-1}\|\mathbf{W}_n\|_F^2\leq P.\label{eq:power_constraint} 
\end{align} \end{subequations} 
The problem in \eqref{eq:original problem} is non-convex due to the matrix inverse in the CRB objective and the nonlinear structure of both the ambiguity and SINR constraints. To achieve a tractable reformulation, we introduce auxiliary variables and rewrite the optimization in terms of transmit covariance matrices.

We introduce an auxiliary positive semidefinite matrix $\mathbf{Q}\in\mathbb{R}^{6\times 6}$ to address the inverse in the CRB objective. Since the FIM, SINR, and ambiguity functions depend on beamformers solely through the covariance matrices  $\mathbf{R}_n=\mathbf{W}_n\mathbf{W}_n^H$ and per-user covariance matrices $\mathbf{R}_{n,k}\triangleq\mathbf{w}_{n,k}\mathbf{w}_{n,k}^H,\forall k$, we equivalently reformulate problem \eqref{eq:original problem} as follows:
\begin{subequations}\label{eq:bf SDP}\begin{align}
&\underset{\mathbf{Q},\mathbf{R}_n,\mathbf{R}_{n,k},\forall k}\min
~~\text{Tr}\{\mathbf{Q}\}\\
&\text{s.t.}\quad \begin{bmatrix}
    \mathbf{F}_\text{gg}   &\bm{\Lambda}^{1/2} \\ \bm{\Lambda}^{1/2} & \mathbf{Q}
\end{bmatrix} \succeq\mathbf{0},\\
& \bigg|\!\sum_{n=0}^{N-1}\text{Tr}\Big\{e^{-\jmath\psi_n(\hat{d}_0)}\sum_{t=1}^{T}\mathbf{a}_{t,n}\mathbf{a}_{t,n}^H \mathbf{R}_n\Big\}\! \bigg| \leq N_\text{t}PT\sqrt{\epsilon_{d_0}},\nonumber \\
&\hspace{6 cm}  \forall \hat{d}_0 \in  \mathcal{D}_\text{side}^{d_0}, \\
&(1\!+\!\Gamma_{n,k}^{-1})\mathbf{h}_{n,k}^H\mathbf{R}_{n,k}\mathbf{h}_{n,k} -\mathbf{h}_{n,k}^H\mathbf{R}_n\mathbf{h}_{n,k}\geq \sigma_\text{c}^2,~\forall n, k, \\
&\sum_{n=0}^{N-1}\text{Tr}\{\mathbf{R}_n\} \leq P,\\
& \mathbf{Q} \in \mathbb{S}_+^{6},\quad \mathbf{R}_n \in \mathbb{S}_+^{N_\text{t}},\quad \mathbf{R}_{n,k} \in \mathbb{S}_+^{N_\text{t}},\ \forall n,k.   \\
&\mathbf{R}_n-\sum\nolimits_{k=1}^K\mathbf{R}_{n,k}\in \mathbb{S}^{N_\text{t}}_+,\\
&\text{Rank}\{\mathbf{R}_{n,k}\}=1,~\forall n, k. \label{eq:rank-one constraint}
\end{align}\end{subequations}
The reformulated problem \eqref{eq:bf SDP} remains non-convex due to the rank-one constraints \eqref{eq:rank-one constraint}. 
To overcome this challenge, we adopt the classical semi-definite relaxation (SDR) technique, temporarily removing the rank-one constraints and converting the problem into a convex semi-definite program (SDP). After solving the relaxed SDP, we obtain the optimal covariance solutions $\widetilde{\mathbf{R}}_n$, $\widetilde{\mathbf{R}}_{n,k}$, from which we subsequently recover the rank-one solutions via \cite{FLiu-TSP-2021}:
\begin{subequations}\label{eq:Rstar}\begin{align}
\mathbf{R}_n^\star &= \widetilde{\mathbf{R}}_n, \\
\mathbf{w}_{n,k}^\star &= (\mathbf{h}_{n,k}^H\widetilde{\mathbf{R}}_{n,k}\mathbf{h}_{n,k})^{-1/2}\widetilde{\mathbf{R}}_{n,k}\mathbf{h}_{n,k},\label{eq:initial wk}\\
\mathbf{R}_{n,k}^\star &=  \mathbf{w}_{n,k}^\star(\mathbf{w}_{n,k}^\star)^H.
\end{align}\end{subequations}
Since $\mathbf{R}_n = \mathbf{W}_n\mathbf{W}_n^H 
= \mathbf{W}_{\text{c},n}\mathbf{W}_{\text{c},n}^H 
+ \mathbf{W}_{\text{s},n}\mathbf{W}_{\text{s},n}^H$, 
we obtain the sensing beamforming covariance as 
\begin{align}
\mathbf{W}_{\text{s},n}\mathbf{W}_{\text{s},n}^H
= \mathbf{R}_n^\star - \sum_{k=1}^K \mathbf{R}_{n,k}^\star.
\end{align}
Thus, $\mathbf{W}_{\text{s},n}^\star$ can be readily obtained via a Cholesky or eigenvalue decomposition.

The dominant computational cost stems from solving the SDP, whose complexity scales as $\mathcal{O}\{N^3K^3N_\text{t}^6\}$ using standard interior-point solvers \cite{FLiu-TSP-2021}. Nevertheless, due to the compact six-dimensional geometric parameterization of the PSM, the proposed optimization framework remains computationally manageable. To highlight the advantages of the proposed PSM framework, we next extend the conventional DSM and UCM approaches to the same MIMO-OFDM ISAC setting, enabling a fair and comprehensive performance comparison.

\section{UCM- and DSM-Based Beamforming Designs}

This section develops beamforming designs based on the UCM and DSM for the MIMO-OFDM ISAC framework established in Sec. III. For each model, we derive the corresponding CRB expressions, formulate the beamforming optimization problems under SINR and power constraints identical to those for the PSM, and analyze their computational complexity. For a fair comparison, all CRBs are mapped into the common PSM parameter space via Jacobian transformations. 


\subsection{UCM-Based Beamforming Optimization}

The UCM represents the extended target response on each subcarrier as an arbitrary unstructured matrix without exploiting any array manifold structure. In the OFDM observation model, the unknown parameters comprise all entries of $\mathbf{G} = [\mathbf{G}_0,\mathbf{G}_1,\dots,\mathbf{G}_{N-1}]$, which yields $N_\text{r}\times N_\text{t}\times N$ complex parameters. The corresponding FIM is given by  
\begin{equation}\label{eq:FG}
\mathbf{F}_{\mathbf{G}} = \frac{L}{\sigma^2_\text{s}}(\mathbf{R}_\text{X}\otimes \mathbf{I}_{N_\text{r}}),   
\end{equation}
where $\mathbf{R}_\text{X}=\text{blkdiag}\{\mathbf{R}_0,\mathbf{R}_1,\dots,\mathbf{R}_{N-1}\}$ is the block-diagonal transmit covariance matrix. Thus, the resulting CRB can be expressed as 
\begin{equation}
\mathbf{C}_{\mathbf{G}} = \mathbf{F}_\mathbf{G}^{-1} = \frac{\sigma_\text{s}^2}{L}(\mathbf{R}_\text{X}^{-1}\otimes\mathbf{I}_{N_\text{r}}),
\end{equation}
and a natural scalar metric for UCM-based sensing is its trace: 
\begin{align}
    \text{Tr}\{\mathbf{C}_{\mathbf{G}}\} = \frac{\sigma_\text{s}^2}{L}\text{Tr}\{\mathbf{R}_\text{X}^{-1}\otimes\mathbf{I}_{N_\text{r}}\} = \frac{\sigma_\text{s}^2N_\text{r}}{L}\sum_{n=0}^{N-1}\text{Tr}\{\mathbf{R}_n^{-1}\}.
\end{align}
The corresponding beamforming optimization is 
\begin{align}
\underset{\mathbf{R}_n}\min~~\sum_{n=0}^{N-1}\text{Tr}\{\mathbf{R}_n^{-1}\},  
\end{align}
subject to the same communication SINR and total power constraints specified in \eqref{eq:SINR_constraint} and \eqref{eq:power_constraint}. Note that due to the absence of geometric parameters in the UCM formulation, ambiguity-function-based sidelobe suppression constraints defined over hypothesized ranges cannot be naturally incorporated. This problem can be solved via the SDR-based optimization procedure described in Sec. III-C. The worst-case computational complexity of this SDP scales as $\mathcal{O}\{N^3K^3N_\text{t}^6\}$, identical to that of the PSM-based beamforming design.

\subsection{DSM-Based Beamforming Optimization}

The DSM models each scatterer individually, resulting in increased parameter dimensionality and computational complexity compared to PSM and UCM. Specifically, the unknown parameter vector in DSM is
\begin{equation}\label{eq:DSM paras}
\begin{aligned}
\bm{\xi}_\text{DSM}&\triangleq\big[\theta_1,\theta_2,\dots,\theta_T,\phi_1,\phi_2,\dots,\phi_T,d_1,d_2,\dots,d_T,\\
&\qquad \Re\{\bm{\alpha}^T\},\Im\{\bm{\alpha}^T\}\big]^T\in\mathbb{R}^{5T}, 
\end{aligned}\end{equation}
which consists of $3T$ geometric parameters (azimuth, elevation, range) and $2T$ real-valued RCS components. The DSM Fisher information matrix $\mathbf{F}^{\text{DSM}} \in \mathbb{R}^{5T\times 5T}$ takes a similar block form as in \eqref{eq:FIM_block}:
\begin{align}\label{FDSM}
\mathbf{F}_{\text{DSM}} =\begin{bmatrix} \mathbf{F}^{\text{DSM}}_\text{gg} & \mathbf{F}^{\text{DSM}}_{\text{g}\alpha}\vspace{0.15 cm} \\ (\mathbf{F}^{\text{DSM}}_{\text{g}\alpha})^T & \mathbf{F}_{\alpha\alpha}^{\text{DSM}} \end{bmatrix} , 
\end{align}
where the geometric information block $\mathbf{F}^{\text{DSM}}_\text{gg}\in\mathbb{R}^{3T\times 3T}$ characterizes the information on individual scatterer locations, and its $(i,j)$-th element is computed as:
\begin{align}
    \mathbf{F}^{\text{DSM}}_\text{gg}(i,j) = \sum_{n=0}^{N-1}\Re\{\text{Tr}\{\mathbf{R}_n\mathbf{C}_{i,j,n}^{\text{DSM}}\}\},
\end{align}
with the expressions for $\mathbf{C}_{i,j,n}^{\text{DSM}}$ derived in Appendix B. Due to the assumption of independence across the scatterers, the matrix $\mathbf{F}^\text{DSM}_\text{gg}$ exhibits a block-sparse structure without inter-scatterer coupling. The corresponding weighted geometric CRB for DSM is formulated as 
$\text{Tr}\{\bm{\Lambda
}_\text{DSM} (\mathbf{F}_\text{gg}^\text{DSM})^{-1}\}$, where $\bm{\Lambda}_\text{DSM}\in\mathbb{R}^{3T\times 3T}$ is a diagonal weighting matrix.

Due to its explicit scatterer-based modeling, DSM requires ambiguity-function-based constraints to be defined separately for each discrete range layer. Letting $\mathcal{T}_r = \{t: d_t=d_r\}$ denote the set of scatterers occupying the $r$-th range bin, the range-specific ambiguity function can be expressed as 
\begin{align}
\chi_{d_r}(\hat{d}_r) = \sigma_\alpha^4N_\text{r}^2 \Big|\sum_{n=0}^{N-1} e^{-\jmath\upsilon_{n,r}(\hat{d}_r)} \sum_{t\in\mathcal{T}_r}\mathbf{a}_{t,n}^H \mathbf{R}_n \mathbf{a}_{t,n}\Big|^2,    
\end{align}
where $\upsilon_{n,r}(\hat{d}_r) = 4\pi n\Delta_f (d_r - \hat{d}_r)/c$ represents the range-induced phase shift. Normalized by the ideal mainlobe amplitude $\chi_{d_r}(d_r)\approx \sigma_\alpha^4N_\text{r}^2N_\text{t}^2P^2T_\theta^2T_\phi^2$, we impose sidelobe suppression constraints as 
\begin{align}
\chi_{d_r}(\hat{d}_r)/\chi_{d_r}(d_r) \leq \epsilon_\text{DSM}, \quad \forall \hat{d}_r \in \mathcal{D}_\text{side}^{(r)},~\forall r,
\end{align}
where $\mathcal{D}_\text{side}^{(r)}$ defines critical sidelobe regions around $d_r$.

Thus, the DSM-based beamforming optimization problem incorporating these ambiguity constraints is formulated as:
\begin{subequations}\begin{align}
&\min_{\mathbf{R}_n} \quad  \text{Tr}\{\bm{\Lambda
}_\text{DSM} (\mathbf{F}_\text{gg}^\text{DSM})^{-1}\} \\
&\text{s.t.} ~ \Big|\sum_{n=0}^{N-1} e^{-\jmath\upsilon_{n,r}(\hat{d}_r)} \sum_{t\in\mathcal{T}_r}\mathbf{a}_{t,n}^H \mathbf{R}_n \mathbf{a}_{t,n}\Big| \leq N_\text{t}PT_\theta T_\phi\sqrt{\epsilon_\text{DSM}},\nonumber \\
&\hspace{5 cm}\forall \hat{d}_r \in \mathcal{D}_\text{side}^{(r)}, ~\forall r, \label{range AF const for DSM}\\
&\qquad\text{\eqref{eq:SINR_constraint}}, ~\text{\eqref{eq:power_constraint}}.\nonumber
\end{align}\end{subequations}
Due to the large dimension ($3T$ geometric parameters) of $\mathbf{F}_\text{gg}^{\text{DSM}}$ and the individually defined ambiguity constraints for each discrete range layer, the DSM-based formulation introduces significantly higher computational complexity compared to the PSM and UCM approaches, and scales as $\mathcal{O}\{N^3K^3N_\text{t}^6 + T^6\}$. This increased complexity significantly restricts the practical applicability of DSM-based designs, particularly for real-time or resource-limited cases.

\subsection{Unified CRB Comparison via Jacobian Transformations}
 
To compare the potential sensing performance achievable using the UCM, DSM, and PSM, we unify their distinct parameterizations into the common geometric parameter space defined by the PSM. This unification is achieved by applying Jacobian transformations to map the respective CRBs for performance comparison without altering the original beamforming solutions obtained under each native model.

For the UCM, we define the Jacobian matrix $\mathbf{J}_{\mathbf{G}}(\bm{\xi})\in\mathbb{C}^{N_\text{r}N_\text{t}N\times (2T+6)}$ that relates the unstructured parameter vector $\text{vec}\{\mathbf{G}\}$ to the structured PSM parameters $\bm{\xi}$: 
\begin{align}
    \mathbf{J}_{\mathbf{G}}(\bm{\xi}) = \frac{\partial\text{vec}\{\mathbf{G}\}}{\partial\bm{\xi}^T},
\end{align}
where the derivatives follow directly from \eqref{eq:Gn} and \eqref{eq:geom pqr}. The Fisher information matrix for UCM transformed into the PSM parameter space is thus expressed as
\begin{align}
    \mathbf{F}_{\bm{\xi}}^\text{UCM} = \mathbb{E}_\alpha\{2\Re\{\mathbf{J}^H_{\mathbf{G}}(\bm{\xi})\mathbf{F}_\mathbf{G}\mathbf{J}_{\mathbf{G}}(\bm{\xi})\}\} + \mathbf{F}_\text{prior},
\end{align}
where $\mathbf{F}_\text{prior}$ contributes solely to the RCS parameter block.

For the DSM, the Jacobian transformation relates the individual scatterer parameters to the compact geometric parameters defined by the PSM. Specifically, the DSM Jacobian matrix $\mathbf{J}_{\text{DSM}}(\bm{\xi}) \in \mathbb{R}^{5T\times(2T+6)}$ is structured as 
\begin{align}
\mathbf{J}_\text{DSM}(\bm{\xi}) = \begin{bmatrix} \mathbf{J}_\text{geo} & \mathbf{0}  \\ \mathbf{0} & \mathbf{I}_{2T} \end{bmatrix}, 
\end{align}
with the geometric block $\mathbf{J}_\text{geo}\in\mathbb{R}^{3T\times 6}$ defined by the partial derivatives:
\begin{equation}\begin{aligned}
\frac{\partial \theta_t}{\partial \theta_0} &= 1, \quad \frac{\partial \theta_t}{\partial \Delta\theta} = u_p, \quad \frac{\partial \theta_t}{\partial [\phi_0, \Delta\phi, d_0, \Delta d]} = 0,\\
\frac{\partial \phi_t}{\partial \phi_0} &= 1, \quad \frac{\partial \phi_t}{\partial \Delta\phi} = v_q, \quad \frac{\partial \phi_t}{\partial [\theta_0, \Delta\theta, d_0, \Delta d]} = 0,\\
\frac{\partial d_t}{\partial d_0} &= 1, \quad \frac{\partial d_t}{\partial \Delta d} = w_r, \quad \frac{\partial d_t}{\partial [\theta_0, \Delta\theta, \phi_0, \Delta\phi]} = 0,
\end{aligned}\end{equation}
with the normalized offsets $u_p, v_q, w_r$ provided in \eqref{eq:geom pqr}. The DSM Fisher information matrix projected onto the PSM parameter space is therefore given by:
\begin{align}
\mathbf{F}_{\xi}^{\text{DSM}} = \mathbb{E}_\alpha\{\mathbf{J}_{\text{DSM}}^T(\bm{\xi}) \mathbf{F}_{\text{DSM}} \mathbf{J}_{\text{DSM}}(\bm{\xi})\} + \mathbf{F}_\text{prior}.    
\end{align}

Under the same assumption as in Sec. III-A, i.e., $\bm{\alpha}\sim\mathcal{CN}(\mathbf{0},\sigma_\alpha^2\mathbf{I}_T)$, the cross-information blocks vanish after expectation with respect to $\bm{\alpha}$. Consequently, the CRBs for the geometric parameters $\bm{\xi}_{\text{geom}}$ under UCM and DSM are obtained by inverting the respective $6\times 6$ leading principal submatrices: 
\begin{subequations}\begin{align}
\mathbf{C}_{\text{geom}}^{\text{UCM}} &= [(\mathbf{F}_{\xi}^{\text{UCM}})_{1:6,1:6}]^{-1},\\
\mathbf{C}_{\text{geom}}^{\text{DSM}} &= [(\mathbf{F}_{\xi}^{\text{DSM}})_{1:6,1:6}]^{-1}.
\end{align}\end{subequations}

These transformed CRBs provide a consistent basis for evaluating and comparing the achievable estimation accuracy across the three scattering models. Furthermore, the Jacobian-based transformations utilized here align with the extended invariance principle (EXIP) \cite{EXIP1}-\cite{EXIP3}, which states that the FIM transforms under parameter changes as $\mathbf{F}_{\bm{\xi}} = \mathbf{J}^T \mathbf{F}_{\bm{\psi}} \mathbf{J}$, given a differentiable reparametrization $\bm{\psi} = \bm{\psi}(\bm{\xi})$. In the subsequent estimation section, we leverage this principle again to show how intermediate estimates produced by UCM and DSM can be optimally mapped onto the compact PSM parameterization via EXIP-based weighted least squares, enabling unified comparisons in the common geometric parameter space.

\section{Maximum A Posteriori Estimation}

Based on the CRB comparisons in Sections III and IV, we now develop MAP estimation algorithms aligned with the proposed PSM and the conventional UCM and DSM approaches. We first establish a unified semi-linear MAP estimation framework that effectively decouples estimation of the nonlinear geometric parameters from linear estimation of the RCS coefficients. Subsequently, we design two-stage EXIP-based estimators for UCM and DSM, mapping their intermediate high-dimensional estimates into the compact geometric PSM parameter space. Finally, we conduct a complexity analysis to highlight the computational advantages of the proposed PSM estimation framework over UCM and DSM.

\subsection{Unified Semi-Linear MAP Estimation}

Consider a generic semi-linear model expressed as:
\begin{equation}
\mathbf{y} = \mathbf{D}(\bm{\zeta}) \bm{\alpha} + \mathbf{n},
\end{equation}
where $\mathbf{y}$ represents the observation, the matrix $\mathbf{D}(\bm{\zeta})$ has a known geometric parameterization $\bm{\zeta}$, and $\bm{\alpha}$ denotes a vector of linear complex coefficients. The noise $\mathbf{n}$ and the parameters $\bm{\alpha}$ are modeled as zero-mean Gaussian random variables:
\begin{equation}
\mathbf{n} \sim \mathcal{CN}(\mathbf{0}, \mathbf{C}_\text{s}),
\qquad
\bm{\alpha} \sim \mathcal{CN}(\mathbf{0}, \mathbf{C}_{\alpha}).
\end{equation}
Under these assumptions, the negative log-posterior is
\begin{equation}\label{eq:joint cost}
J(\bm{\zeta}, \bm{\alpha}) 
= 
\big\| \mathbf{C}_\text{s}^{-\frac{1}{2}}
\bigl( \mathbf{y} - \mathbf{D}(\bm{\zeta}) \bm{\alpha} \bigr)
\big\|_2^2
+
\big\| \mathbf{C}_{\alpha}^{-\frac{1}{2}} \bm{\alpha} \big\|_2^2 .
\end{equation}
Exploiting the linearity with respect to $\bm{\alpha}$, we first analytically minimize \eqref{eq:joint cost} over $\bm{\alpha}$ to obtain the closed-form MAP estimate:
\begin{equation}\label{eq:alpha_MAP}
\hat{\bm{\alpha}}(\bm{\zeta})
=\left[
\mathbf{D}^H(\bm{\zeta}) \mathbf{C}_\text{s}^{-1} \mathbf{D}(\bm{\zeta})
+
\mathbf{C}_{\alpha}^{-1}
\right]^{-1}
\mathbf{D}^H(\bm{\zeta}) \mathbf{C}_\text{s}^{-1} \mathbf{y}.
\end{equation}
Substituting \eqref{eq:alpha_MAP} into \eqref{eq:joint cost} and omitting constants independent of $\bm{\zeta}$, we derive the resulting MAP estimator for $\zeta$:
\begin{subequations}\begin{align}
\hat{\bm{\zeta}} &= \arg \min_{\bm{\zeta}} \; J(\bm{\zeta}) \\
&= \arg \min_{\bm{\zeta}} \; -\mathbf{p}^H(\bm{\zeta}) \mathbf{B}^{-1}(\bm{\zeta}) \mathbf{p}(\bm{\zeta}), 
\label{eq:geom_optimization} 
\end{align}\end{subequations}
\begin{subequations} \label{eq:Bp}\begin{align}
\mathbf{B}(\bm{\zeta}) &= \mathbf{D}^H(\bm{\zeta}) \mathbf{C}_\text{s}^{-1} \mathbf{D}(\bm{\zeta}) + \mathbf{C}_{\alpha}^{-1}, \\
\mathbf{p}(\bm{\zeta}) &= \mathbf{D}^H(\bm{\zeta}) \mathbf{C}_\text{s}^{-1} \mathbf{y}.
\end{align}\end{subequations}

To efficiently solve \eqref{eq:geom_optimization}, we derive the gradient of $J(\bm{\zeta})$. Using the variable projection theorem, the gradient with respect to each geometric parameter $\zeta_i$ can be expressed as
\begin{equation}\label{eq:J derivative}
\frac{\partial J(\bm{\zeta})}{\partial \zeta_i}
= -2 \Re \left\{
\mathbf{p}^H\mathbf{B}^{-1}
\frac{\partial \mathbf{D}^{H}}{\partial \zeta_i}
\mathbf{C}_\text{s}^{-1}(\mathbf{y}-\mathbf{DB}^{-1}\mathbf{p})
\right\}.
\end{equation}
Subsequently, $\bm{\zeta}$ is iteratively updated via gradient-based optimization combined with the Armijo backtracking line search. Given the current estimate $\bm{\zeta}^{(m)}$, the update is given by
\begin{equation}\label{eq:geom_update}
\bm{\zeta}^{(m+1)} = \bm{\zeta}^{(m)} - \mu^{(m)} \nabla J(\bm{\zeta}^{(m)}),
\end{equation}
where the adaptive step size $\mu^{(m)}$ is determined by the Armijo
rule to guarantee sufficient descent and robust convergence. Upon convergence, the linear coefficients $\bm{\alpha}$ are recovered using \eqref{eq:alpha_MAP} evaluated with $\hat{\bm{\zeta}}$.

For the proposed PSM, we have $\bm{\zeta} = \bm{\xi}_\text{geom} = [
\theta_0,\,\Delta \theta,\,\phi_0,\,\Delta \phi,\,d_0,\,\Delta d
]^T$, $\mathbf{y} = \operatorname{vec}(\mathbf{Y})$ from~\eqref{eq:observe Y}, and $\mathbf{D}(\bm{\xi}_\text{geom})$ is formed from the PSM steering matrices defined in \eqref{eq:Dxi}. 
With these definitions, the PSM estimation procedure is detailed in
Algorithm~1. In contrast, MAP estimation for UCM and DSM inherently requires two sequential estimation stages due to their initial step of estimating high-dimensional intermediate parameters. For UCM, one must initially estimate per-subcarrier unstructured channel matrices, while DSM involves initial nonlinear estimation of the individual scatterer-level parameters. Both methods necessitate a second estimation step that maps their intermediate high-dimensional outputs to the compact geometric PSM representation. This structured parameter extraction, which exploits the intrinsic geometric constraints and asymptotic Gaussian properties of the intermediate estimates, can be elegantly implemented using the EXIP-based procedure described next.

\begin{algorithm}[!t]
    \begin{small}
        \caption{Semi-Linear MAP Estimation for PSM}
        \label{alg:psm_mle}
        \begin{algorithmic}[1]
            \REQUIRE {Observation matrix $\mathbf{Y}$, transmit signal matrix $\mathbf{X}$, noise variance $\sigma_\text{s}^2$, prior variance $\sigma_\alpha^2$, initial geometric parameters $\bm{\xi}_{\text{geom}}^{(0)}$, tolerance $\varepsilon_\text{tol}$, maximum iterations $K_{\text{max}}$.}
            \ENSURE {Geometric parameters $\hat{\bm{\xi}}_{\text{geom}}$ and RCS coefficients $\hat{\bm{\alpha}}$.}
            \STATE {Initialize $m=0$, form the observation vector $\mathbf{y}=\text{vec}\{\mathbf{Y}\}$.}     
            \REPEAT
            \STATE {Construct dictionary matrix $\mathbf{D}(\bm{\xi}_{\text{geom}}^{(m)})$.}            
            \STATE {Compute $\mathbf{B}^{(m)}$ and $\mathbf{p}^{(m)}$ via \eqref{eq:Bp}.}
            \STATE {Compute gradient vector $\nabla \mathcal{J}(\bm{\xi}_{\text{geom}}^{(m)})$ via \eqref{eq:J derivative}.}
            \STATE {Determine step size $\mu^{(m)}$ via Armijo backtracking line search.}
            \STATE {Update geometric parameters by \eqref{eq:geom_update}.}
            \STATE {Set $m \leftarrow m+1$.}
            \UNTIL {$\|\bm{\xi}_\text{geom}^{(m)}-\bm{\xi}_\text{geom}^{(m-1)}\|^2\leq \varepsilon_\text{tol}$ or $m \geq K_{\max}$.}
            \STATE {Return $\hat{\bm{\xi}}_{\text{geom}} = \bm{\xi}_{\text{geom}}^{(m)}$.}
            \STATE {Compute RCS estimates using \eqref{eq:alpha_MAP} at $\hat{\bm{\xi}}_{\text{geom}}$.}
        \end{algorithmic}
    \end{small}
\end{algorithm}

\subsection{Structured Parameter Estimation via EXIP}

EXIP provides a rigorous way to extract structured parameter estimates using intermediate higher-dimensional models \cite{EXIP1,EXIP2,EXIP3}. 
Suppose an intermediate estimate $\hat{\bm{\psi}}$ is unbiased with an asymptotic Gaussian distribution: $\hat{\bm{\psi}}-\bm{\psi} \sim\mathcal{CN}(\mathbf{0},\mathbf{C}_\psi)$, where the true parameters $\bm{\psi}$ obey a known differentiable structural mapping $\bm{\psi} = \bm{\psi}(\bm{\zeta})$. According to EXIP, an asymptotically efficient estimator for the structured parameters $\bm{\zeta}$ can be obtained by solving the following weighted least-squares (WLS) optimization:
\begin{equation}
\hat{\bm{\zeta}} = \arg \min_{\bm{\zeta}}\left\|
\mathbf{C}_{\psi}^{-\frac{1}{2}}\bigl(\hat{\bm{\psi}} - \bm{\psi}(\bm{\zeta})
\bigr)\right\|_2^2,
\end{equation}
possibly incorporating an additional prior or regularization term. This EXIP approach can be applied to MAP estimates obtained using the UCM and DSM approaches, since in both cases the estimation error is asymptotically Gaussian with zero mean and covariance given by the associated CRBs derived in Section IV-C. We detail this approach in the next two sections.


\subsection{UCM Estimation Using EXIP}

For UCM, the MAP estimate of the channel response at each subcarrier $n$ is given by the least-squares solution
\begin{align}\label{eq:Gn estimate}
\hat{\mathbf{G}}^{\text{UCM}}_n = \mathbf{Y}_n\mathbf{X}_n^H(\mathbf{X}_n\mathbf{X}_n^H)^{-1},\quad n=0,\dots,N-1, 
\end{align}
assuming that $(\mathbf{X}_n\mathbf{X}_n^{H})^{-1}$ exists. 
The channel estimates are then concatenated into a single high-dimensional vector:
\begin{align}
\hat{\bm{\psi}} = \big[\text{vec}^{T}\{\hat{\mathbf{G}}^{\text{UCM}}_{0}\},
\ldots, \text{vec}^{T}\{\hat{\mathbf{G}}^{\text{UCM}}_{N-1}\}
\big]^{T}.
\end{align}

In the second stage, we map the unstructured intermediate estimate $\hat{\bm{\psi}}$ onto the compact PSM geometric parameter space via EXIP, as follows:
\begin{subequations}\begin{align}
\bm{\psi}(\bm{\xi}_\text{geom},\bm{\alpha}) &= \big[
\text{vec}^T\{\mathbf{G}_0\}, \dots, \text{vec}^T\{\mathbf{G}_{N-1}\}
\big]^{T}\\
& \triangleq \mathbf{M}(\bm{\xi}_\text{geom})\bm{\alpha},         
\end{align}\end{subequations}
where the matrix $\mathbf{M}(\bm{\xi}_\text{geom})\in\mathbb{C}^{NN_\text{r}N_\text{t}\times T}$ has  columns: 
\begin{equation}
\mathbf{m}_t(\bm{\xi}_\text{geom})\triangleq
\big[\text{vec}\{\mathbf{V}_{t,0}\}^T,\ldots,
\text{vec}\{\mathbf{V}_{t,N-1}\}^T\big]^T  ,  
\end{equation}
and $\mathbf{V}_{t,n}(\bm{\xi}_\text{geom})=f_{t,n}\mathbf{b}_{t,n}\mathbf{a}_{t,n}^H$ as defined in Sec.~II.
The resulting EXIP-based MAP estimator is thus given by:
\begin{equation}\label{eq:ucm_exip_wls} 
\underset{\bm{\xi}_{\text{geom}},\bm{\alpha}}{\min}~~
\left\|\mathbf{F}_{\mathbf{G}}^{\frac{1}{2}}\left(\bm{\hat{\psi}}-\bm{\psi}(\bm{\xi}_\text{geom},\bm{\alpha})\right)\right\|^2 + \frac{1}{\sigma_\alpha^2}\|\bm{\alpha}\|^2,
\end{equation}
where $\mathbf{F}_{\mathbf{G}}$ is the FIM for UCM defined in~\eqref{eq:FG}.
Given fixed geometric parameters $\bm{\xi}_\text{geom}$, the above optimization simplifies to a linear ridge regression problem for the RCS coefficients $\bm{\alpha}$. Substituting the optimal $\bm{\alpha}$ solution back into \eqref{eq:ucm_exip_wls} yields a nonlinear optimization problem with respect to the six geometric parameters in $\bm{\xi}_\text{geom}$. This reduced-dimensional problem is efficiently solved using Algorithm~1 with $\mathbf{y} = \hat{\bm\psi}$, $\mathbf{D} = \mathbf{M}(\bm{\xi}_\text{geom})$, $\mathbf{C}_\text{s} = \mathbf{F}_\mathbf{G}^{-1}$, and $\mathbf{C}_\alpha = \sigma_\alpha^2\mathbf{I}$.

\subsection{DSM Estimation Using EXIP}
 
Under DSM, the unknown parameters consist of individual scatterer coordinates and their associated RCS coefficients, i.e., 
$\{\theta_t,\phi_t,d_t,\alpha_t\}_{t=1}^T$. The corresponding MAP estimation problem is formulated as
\begin{equation}
\min_{\{\theta_t, \phi_t, d_t, \alpha_t\}_{t=1}^T} \frac{1}{\sigma_\text{s}^2}\big\|\mathbf{Y} - \sum_{t=1}^T \alpha_t\mathbf{V}_t(\theta_t, \phi_t, d_t)\mathbf{X}\big\|_F^2 + \frac{1}{\sigma_\alpha^2}\|\bm{\alpha}\|^2. 
\end{equation}
Following the unified MAP estimation framework in Section V-A, we first eliminate the linear parameters $\bm{\alpha}$ via a closed-form solution and subsequently perform a gradient-based nonlinear search over the remaining $3T$-dimensional scatterer coordinate vector. Detailed expressions for the required steering-vector derivatives are provided in Appendix~B. 

The scatterer-wise coordinate estimates obtained from the first stage are stacked into the intermediate parameter vector: $\hat{\bm{\vartheta}} = [\hat{\theta}_1,\dots,\hat{\theta}_T,\hat{\phi}_1,\dots,\hat{\phi}_T,\hat{d}_1,\dots,\hat{d}_T]^T\in\mathbb{R}^{3T}$. 
These parameters are linearly related to the compact geometric parameter vector via $\bm{\vartheta} = \mathbf{U}\bm{\xi}_\text{geom}$, where 
\begin{align}
\mathbf{U} = \begin{bmatrix}
\mathbf{1}_T & \mathbf{u} & \mathbf{0} & \mathbf{0} & \mathbf{0} & \mathbf{0} \\
\mathbf{0} & \mathbf{0} & \mathbf{1}_T & \mathbf{v} & \mathbf{0} & \mathbf{0} \\
\mathbf{0} & \mathbf{0} & \mathbf{0} & \mathbf{0} & \mathbf{1}_T & \mathbf{w}
\end{bmatrix} \in\mathbb{R}^{3T\times 6} ,     
\end{align}
and $\mathbf{u}$, $\mathbf{v}$, $\mathbf{w}$ contain the normalized offsets defined in \eqref{eq:geom pqr}.
Let $\hat{\bm{\alpha}}$ denote the first-stage estimate of the RCS coefficients. The complete vector of intermediate DSM parameter estimates is then given by $\hat{\bm{\xi}}_\text{DSM} = [\hat{\bm{\vartheta}}^T~\Re\{\hat{\bm{\alpha}}\}^T~\Im\{\hat{\bm{\alpha}}\}^T]^T$. The corresponding underlying parameter vector modeled using $\bm{\xi}_\text{geom}$ can be written as $\bm{\xi}_\text{DSM} = [(\mathbf{U}\bm{\xi}_\text{geom})^T~\Re\{\bm{\alpha}\}^T~\Im\{\bm{\alpha}\}^T]^T$.

The Stage-1 estimation error is asymptotically zero-mean Gaussian with covariance given by the inverse of $\mathbf{F}_\text{DSM}$ defined in~\eqref{FDSM}. 
Following EXIP, the second-stage structured fit minimizes a weighted least-squares criterion. Incorporating the Gaussian prior on $\bm{\alpha}$, the joint EXIP estimator is given by 
\begin{equation} 
\underset{\bm{\xi}_{\text{geom}},\bm{\alpha}}{\min}~~
\big\|\mathbf{F}_\text{DSM}^{\frac{1}{2}}(\hat{\bm{\xi}}_\text{DSM}-\bm{\xi}_\text{DSM} )\big\|^2 + \frac{1}{\sigma_\alpha^2}\|\bm{\alpha}\|^2.
\end{equation}
Under the adopted hybrid Bayesian model where $\bm{\alpha}\sim\mathcal{CN}(\mathbf{0},\sigma_\alpha^2\mathbf{I}_T)$, the expected geometric--RCS cross-information vanishes, i.e., $\mathbf{F}_{\text{g}\alpha}^\text{DSM}=\mathbf{0}$, and hence the effective geometric information reduces to the DSM geometric FIM block $\mathbf{F}_\text{gg}^\text{DSM}$, enabling the EXIP refinement to separately update the geometric and the RCS parameters via two independent subproblems:
\begin{align}
\hat{\bm{\xi}}_{\text{geom}}^{\text{DSM}} &= \arg\min_{\bm{\xi}_{\text{geom}}} \big\|(\mathbf{F}_\text{gg}^\text{DSM})^{\frac{1}{2}}(\hat{\bm{\vartheta}}- \mathbf{U}\bm{\xi}_{\text{geom}})\big\|^2,\\
\hat{\bm{\alpha}}^{\text{DSM}} &= \arg\min_{\bm{\alpha}} \big\|(\mathbf{F}_{\alpha\alpha}^\text{DSM})^{\frac{1}{2}}(\hat{\bm{\alpha}}- \bm{\alpha})\big\|^2 + \frac{1}{\sigma_\alpha^2}\|\bm{\alpha}\|^2,
\end{align}
which admit the following closed-form solutions:
\begin{align}\label{eq:DSM to PSM}
\hat{\bm{\xi}}_{\text{geom}}^{\text{DSM}} & = (\mathbf{U}^T\mathbf{F}_\text{gg}^\text{DSM}\mathbf{U})^{-1}\mathbf{U}^T\mathbf{F}_\text{gg}^\text{DSM}\hat{\bm{\vartheta}},\\
\hat{\bm{\alpha}}^{\text{DSM}} &=   (\mathbf{F}_{\alpha\alpha}^\text{DSM}+\frac{1}{\sigma_\alpha^2}\mathbf{I}_{2T})^{-1}\mathbf{F}_{\alpha\alpha}^\text{DSM}\hat{\bm{\alpha}}.
\end{align}
This direct, closed-form EXIP procedure efficiently projects the high-dimensional intermediate DSM estimates into the compact geometric subspace defined by the PSM.

\subsection{Complexity Comparison}
Although the MAP estimators for the UCM, DSM and PSM all share a similar semi-linear variable-projection structure, their complexities differ markedly due to the dimensionality of their respective nonlinear optimization problems. The proposed PSM approach significantly reduces complexity by constraining the nonlinear search to six geometric parameters, striking a favorable balance between computational efficiency and modeling flexibility compared to the DSM and UCM approaches. This will be demonstrated in the following section.

\section{Simulation Results} 

In this section, we present simulation results to evaluate the performance of beamforming optimization for the PSM, UCM, and DSM models and their associated MAP estimation algorithms for extended-target sensing in MIMO-OFDM ISAC systems. In addition, a radar-only benchmark obtained by removing the communication SINR constraints is included to characterize the fundamental limit of sensing performance. 
Unless otherwise specified, we consider a system configured with $N_\text{tx}=N_\text{ty} = 4$ transmit antennas and $N_\text{rx}=N_\text{ry} = 36$ receive antennas, operating at a carrier frequency of $f_\text{c} = 28$GHz with subcarrier spacing $\Delta_f = 480$kHz. Each CPI contains $L=32$ OFDM symbols and $N=128$ subcarriers, simultaneously serving $K=6$ communication users with an SINR requirement of $\Gamma_{n,k} = 10$dB. The receiver noise power is $-90$dBm. The users are located $50$m away from the BS with a path-loss exponent of $3$, and their channels follow a Rician fading model with a Rician factor of $10$. For sensing, an extended target is located at $d_0 = 25$m with a path-loss exponent of $2.6$. The target occupies a 3D region of $12 \text{m} \times 2.5 \text{m} \times 3.8 \text{m}$ and is rotated by $45^\circ$ relative to the BS, resulting in azimuth, elevation, and range extents of $\Delta_\theta = 23.50^\circ$, $\Delta_\phi = 8.71^\circ$, and $\Delta_d = 10.25$m, respectively. Discretization points are selected for $T_\theta = 4$, $T_\phi=2$, and $T_d = 3$. The thresholds for range sidelobe levels are set to $\epsilon_{d_0} = \epsilon_\text{DSM} = 10^{-2}$, and the sidelobe control region is defined over ten range bins surrounding the actual target range. 

\begin{figure}[!t]
    \centering
    \includegraphics[width=\linewidth]{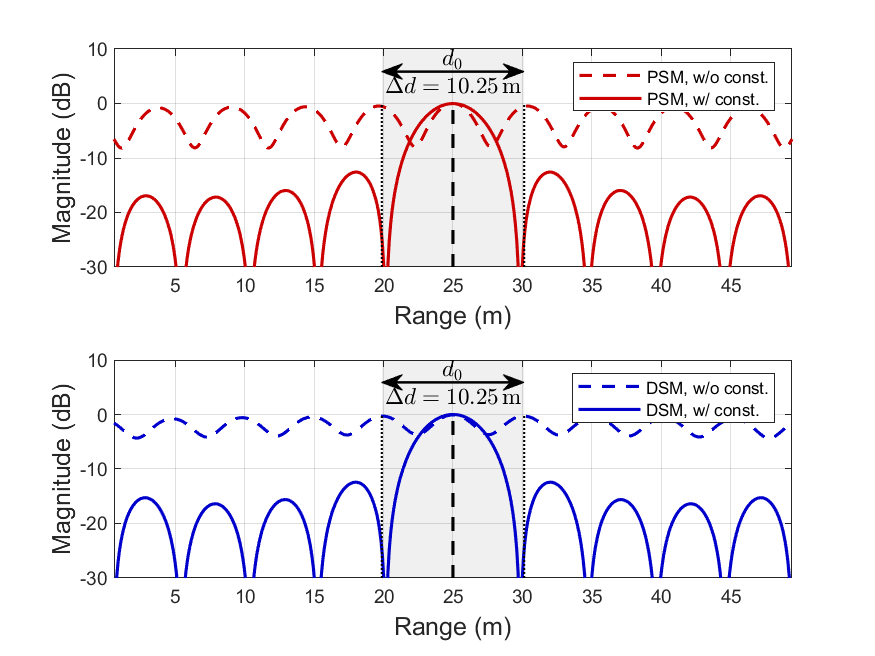}
    \caption{Range ambiguity function.}\vspace{-0.2 cm}
    \label{fig:AF}
\end{figure}

We first examine the range ambiguity functions in Fig.~\ref{fig:AF}, comparing the PSM and DSM designs with and without the proposed range sidelobe suppression constraints (\eqref{eq:ambiguity_constraint} for PSM and \eqref{range AF const for DSM} for DSM). As discussed in Sec.~III, CRB-driven designs without sidelobe constraints produce pronounced periodic sidelobes, e.g., around 20m and 30m in Fig.~\ref{fig:AF}, potentially causing range ambiguity. By incorporating the proposed constraints, these periodic sidelobes are suppressed by more than 30 dB, significantly improving the ambiguity profile and enhancing range estimation robustness.

\begin{figure}[!t]
    \centering
    \includegraphics[width=\linewidth]{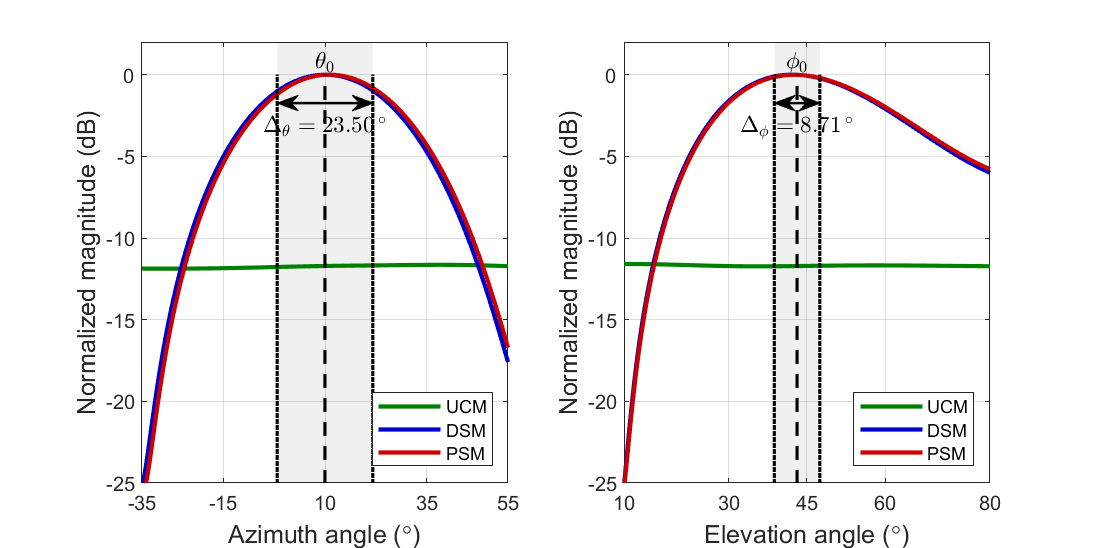}
    \caption{Transmit spatial beampattern.}
    \vspace{-0.4 cm}
    \label{fig:beampattern}
\end{figure}

To illustrate the differences among the three CRB-based beamforming formulations, Fig.~\ref{fig:beampattern} presents their respective optimized transmit beampatterns. Specifically, the PSM and DSM methods leverage target geometric priors along with the array manifold structure, thus generating spatially focused radiation patterns matched to the angular support of the extended target. In contrast, the UCM method, lacking geometry-informed constraints, yields a broader and nearly omnidirectional power distribution, inevitably resulting in lower effective power delivered to the target region. These beampattern discrepancies help explain the sensing performance differences observed in subsequent numerical evaluations.

\begin{figure}[!t] 
\centering
    \includegraphics[width=\linewidth]{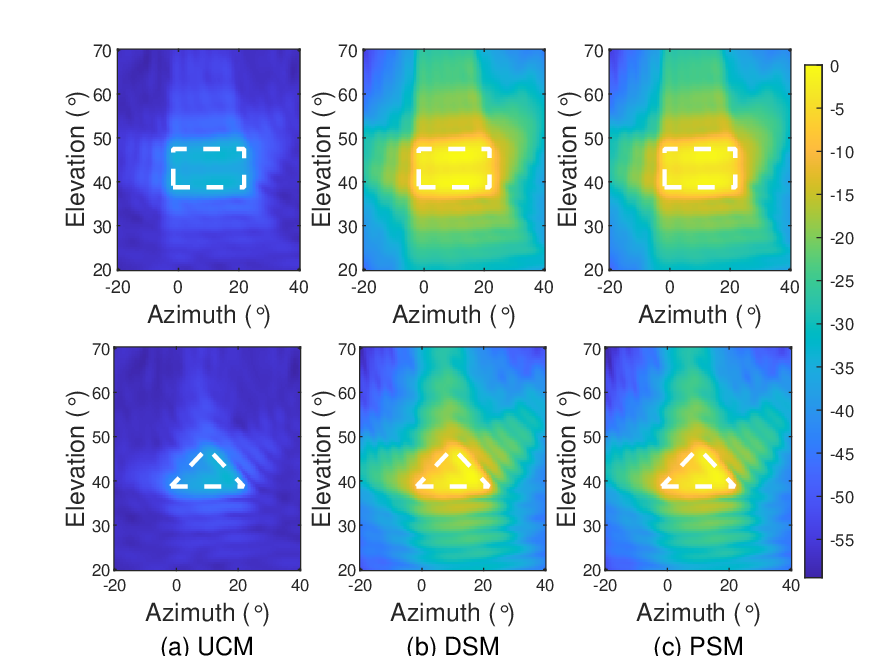}
     \caption{Recovered angular profiles.}
     \vspace{-0.4 cm}
     \label{fig:heatmap}
 \end{figure}
 
Fig. \ref{fig:heatmap} shows the recovered angular-domain target profiles for two representative shapes, with the true boundaries marked by white dashed lines. The RCS is set to be nonzero only within the boundaries and zero elsewhere. Both the proposed PSM and the conventional DSM methods achieve focused energy distributions well-aligned with the actual target extent, effectively highlighting the correct shape and clearly differentiating the targets from the surrounding region. In contrast, the UCM approach produces diffuse angular images with weak target echoes that are nearly obscured by background noise. 

\begin{figure}[!t]
    \centering
    \includegraphics[width=\linewidth]{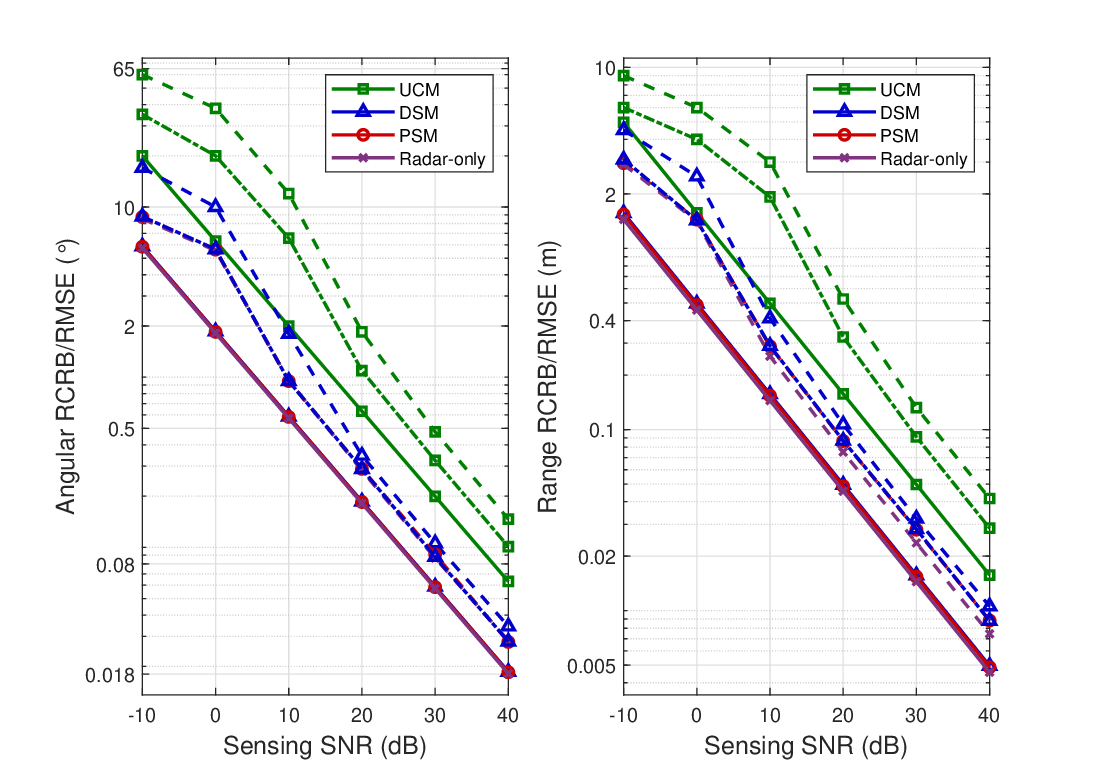}
    \caption{RCRB/RMSE versus sensing SNR. }\vspace{-0.4 cm}
    \label{fig:power}
\end{figure}

The next three plots show performance in the PSM geometric parameter space, evaluating the summed root-CRBs (RCRBs) and RMSEs for the four angular parameters $(\theta_0, \Delta_\theta, \phi_0, \Delta_\phi)$ and two range parameters $(d_0,\Delta_d)$. For fair comparison, the CRBs derived under the UCM and DSM are mapped into the PSM space via Jacobian transformations, as derived in Section IV-C. In Figs. 4-6, \textbf{solid lines} indicate the RCRBs, and \textbf{dashed lines} represent RMSEs obtained using each model's matched estimator. The PSM and radar-only benchmarks use the direct semi-linear MAP estimator, while DSM and UCM employ their respective two-stage EXIP implementations. In addition, to highlight the advantage of direct geometric estimation, we present RMSE curves (\textbf{dash-dotted lines}) obtained by applying the direct PSM MAP estimator to the DSM- and UCM-optimized beamformers.

\begin{figure}[!t]
    \centering
    \includegraphics[width=\linewidth]{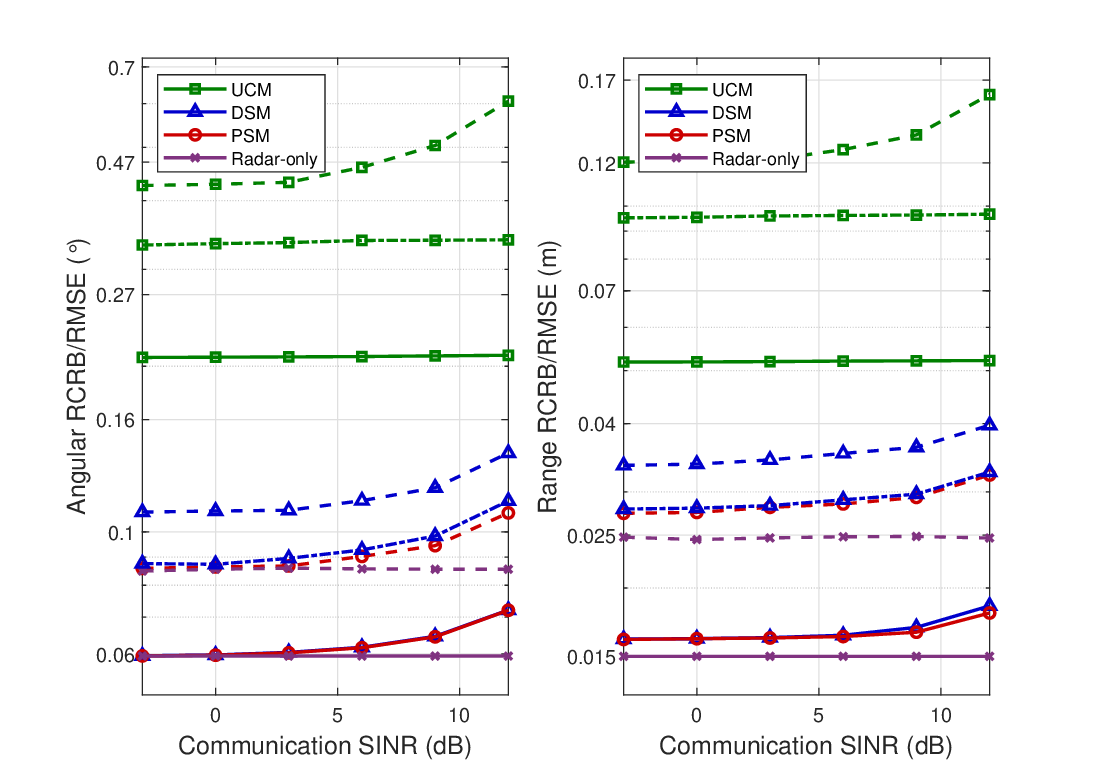}
    \caption{RCRB/RMSE versus SINR requirements.} \vspace{-0.4 cm}
    \label{fig:SINR}
\end{figure}

Fig. \ref{fig:power} presents the angle and range estimation performance as functions of the sensing SNR, defined by the approximate received sensing signal-to-noise ratio: $PLN_\text{r}N_\text{t}\sigma_\alpha^2/\sigma_\text{s}^2$. 
At the theoretical level, the proposed PSM-based design achieves RCRB performance close to the DSM and radar-only benchmarks and significantly outperforms the UCM approach. This demonstrates that the compact geometric representation employed by the PSM is sufficient to attain the same fundamental sensing performance limits as the more complex DSM formulation, despite its substantially reduced model dimensionality. At the estimator level, the proposed PSM-based MAP estimator closely matches the corresponding RCRB, especially at high SNR, and consistently outperforms the two-stage EXIP estimators employed by DSM and UCM. The superior performance of the PSM estimator arises from its direct, low-dimensional parameter estimation, which avoids error propagation inherent in the two-stage methods.

Fig. \ref{fig:SINR} quantifies the fundamental sensing-communication trade-off as a function of the per-user communication SINR requirement. As the SINR constraints become more stringent, more spatial degrees of freedom and power resources must be dedicated to satisfying the users' QoS, thereby reducing flexibility for sensing optimization and causing the sensing RCRB/RMSE to gradually degrade.

\begin{figure}[!t]
    \centering
    \includegraphics[width=\linewidth]{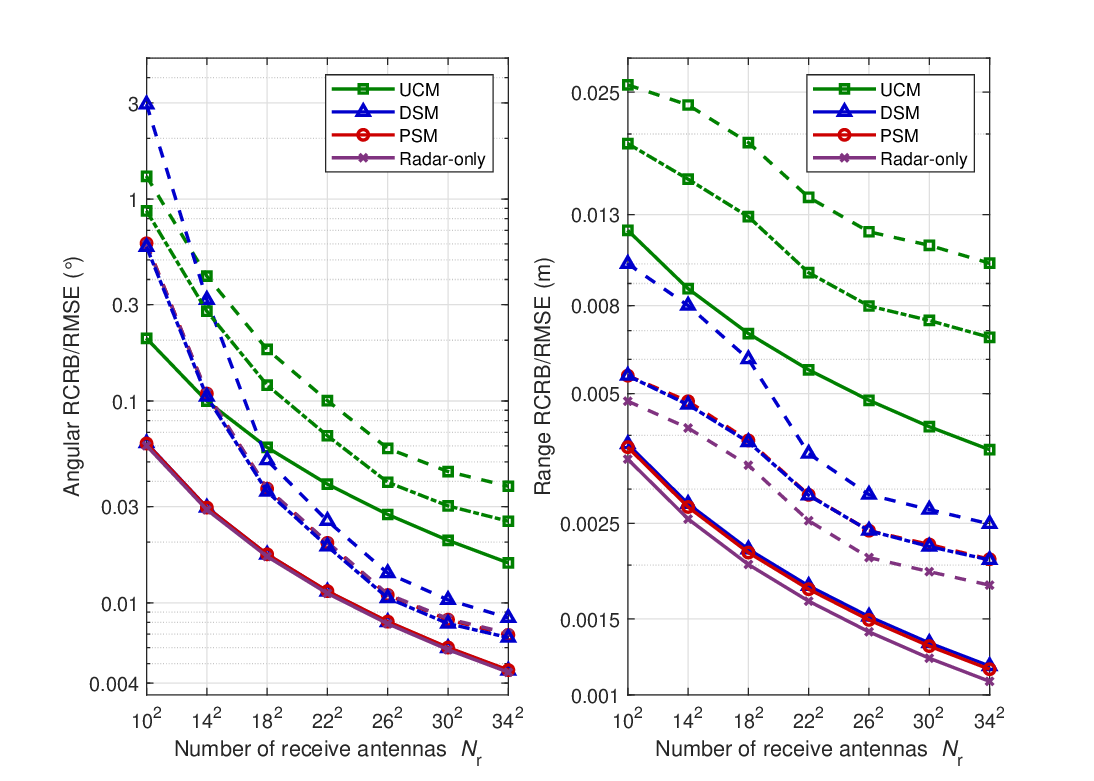}
    \caption{RCRB/RMSE versus number of receive antennas.}\vspace{-0.4 cm}
    \label{fig:Nr}
\end{figure}

Fig. \ref{fig:Nr} evaluates the sensing performance versus the number of receive antennas (with $N_\text{rx} = N_\text{ry}$). As the receive array size grows, both angular and range estimation errors decrease, due to the enhanced spatial resolution and array gain. In addition, the gap between the RMSE and its corresponding RCRB progressively shrinks as the array size increases, indicating that the estimators become more efficient and accurate with larger antenna arrays.



\begin{figure}[!t]
    \centering
    \includegraphics[width=\linewidth]{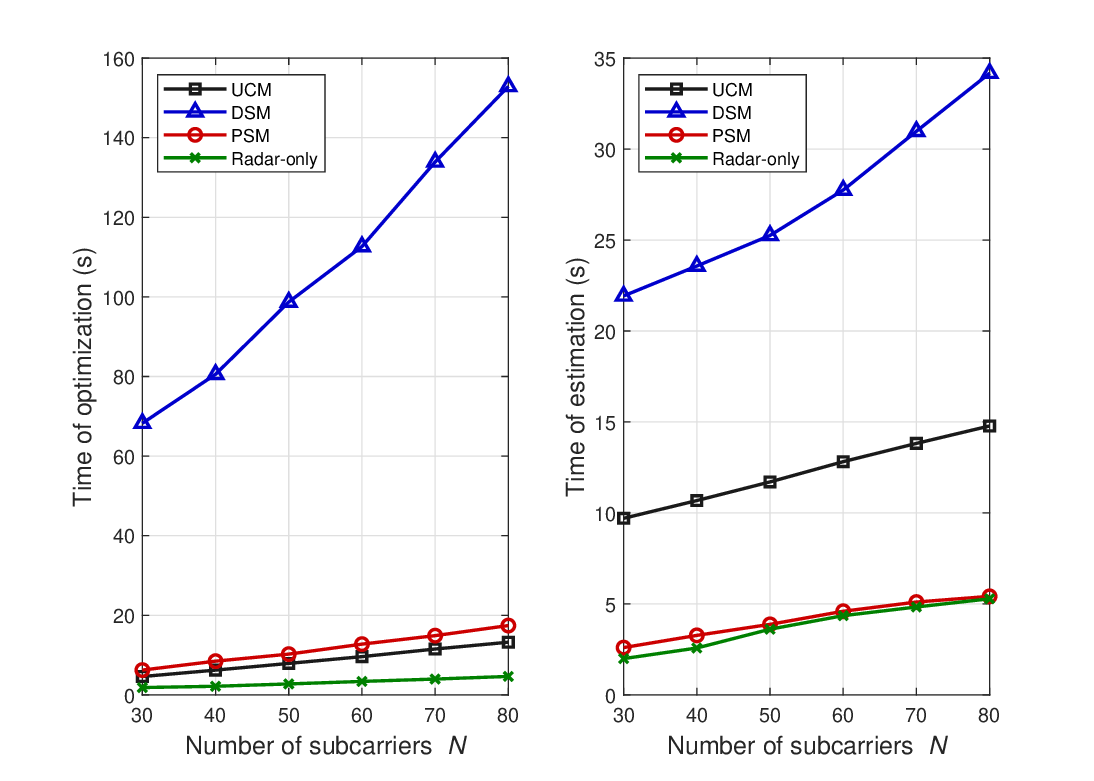}
    \caption{Execution time versus number of subcarriers $N$.}\vspace{-0.4 cm}
    \label{fig:runtime}
\end{figure}

Finally, Fig. \ref{fig:runtime} compares the execution runtime of the beamforming optimization and parameter estimation for different scattering models. The simulations are performed in MATLAB R2023b on a computer equipped with an Apple M4 Pro processor and 24 GB of memory, using CVX with the MOSEK solver. The results indicate that the proposed PSM framework achieves substantial computational savings compared to UCM and DSM, reducing the beamforming optimization and parameter estimation times by $89\%$ and $85\%$, respectively, compared with the DSM. The UCM achieves a larger average reduction in optimization time but a smaller reduction in estimation time since it employs a high-dimensional unstructured channel matrix in the initial step. Although the DSM offers higher modeling fidelity, its large number of parameters substantially increases the complexity and nonlinearity of both optimization and estimation. Combined with the sensing-accuracy results, these runtime comparisons highlight that the proposed PSM strikes a more favorable performance-complexity trade-off, achieving comparable estimation accuracy with markedly lower computational overhead.

\section{Conclusions}

We have proposed a unified modeling, optimization, and estimation framework for extended-target sensing in MIMO-OFDM ISAC systems. We introduced a parametric scattering model that compactly represents extended targets via a six-dimensional geometric parameterization, significantly reducing complexity compared to conventional point cloud (DSM) and unstructured (UCM) frameworks while preserving modeling flexibility and interpretability. Leveraging this model, we derived the Bayesian CRB for joint angle-range estimation and designed CRB-oriented beamformers with range ambiguity suppression constraints to address intrinsic OFDM ambiguities. Furthermore, we proposed a computationally efficient semi-linear MAP estimator tailored specifically to the compact geometry of the PSM, avoiding the complexity of the high-dimensional intermediate estimations required by DSM and UCM. Simulation results demonstrated that the PSM approach achieves estimation performance comparable to the more complex DSM-based designs and radar-only benchmarks, while significantly outperforming the UCM method. The PSM also offers substantial computational advantages, highlighting its strong practical potential for ISAC applications.

\begin{appendices}
\section{FIM Coefficient Matrices under PSM}

We derive the matrices $\mathbf{C}_{i,j,n}$ required for calculation of the FIM block $\mathbf{F}_\text{gg}$ in Sec. III-A. Under the PSM described in Sec. II-D, the extended target response matrix is  
\begin{align}
    \mathbf{G}(\bm{\xi}) = \sum_{t=1}^T\alpha_t\mathbf{V}_t,
\end{align}
where each scatterer contributes via $\mathbf{V}_t = [\mathbf{V}_{t,0},\dots,\mathbf{V}_{t,N-1}]$ with $\mathbf{V}_{t,n} = f_{t,n}\mathbf{b}_{t,n}\mathbf{a}_{t,n}^H$. 
The derivatives of $\mathbf{V}_{t,n}$ w.r.t. the geometric parameters $\bm{\xi}_\text{geom}$ follow from the chain rule. 
For center parameters $\theta_0$, $\phi_0$, $d_0$, these derivatives are given by
\begin{subequations}\label{eq:center-derivatives}\begin{align}
\dot{\mathbf{V}}_{t,n,1}&\triangleq \frac{\partial\mathbf{V}_{t,n}}{\partial\theta_0} =  f_{t,n}\frac{\partial\mathbf{b}_{t,n}}{\partial\theta_0}\mathbf{a}_{t,n}^H + f_{t,n}\mathbf{b}_{t,n}\frac{\partial^H\mathbf{a}_{t,n}}{\partial\theta_0},\\
\dot{\mathbf{V}}_{t,n,3}&\triangleq\frac{\partial\mathbf{V}_{t,n}}{\partial\phi_0}  = f_{t,n}\frac{\partial\mathbf{b}_{t,n}}{\partial\phi_0}\mathbf{a}_{t,n}^H + f_{t,n}\mathbf{b}_{t,n}\frac{\partial^H\mathbf{a}_{t,n}}{\partial\phi_0},\\
\dot{\mathbf{V}}_{t,n,5}&\triangleq\frac{\partial\mathbf{V}_{t,n}}{\partial d_0}  = \frac{\partial f_{t,n}}{\partial d_0}\mathbf{b}_{t,n}\mathbf{a}_{t,n}^H.
\end{align}\end{subequations}

For extent parameters $\Delta_\theta$, $\Delta_\phi$, $\Delta_d$, the derivatives directly scale the corresponding center-parameter derivatives, reflecting the spatial offsets defined in Sec. II-D:
\begin{subequations}\label{eq:extent-derivatives}\begin{align}
\dot{\mathbf{V}}_{t,n,2}&\triangleq\frac{\partial\mathbf{V}_{t,n}}{\partial\Delta_\theta} = u_t\dot{\mathbf{V}}_{t,n,1} ,\\
\dot{\mathbf{V}}_{t,n,4}&\triangleq\frac{\partial\mathbf{V}_{t,n}}{\partial\Delta_\phi} = v_t\dot{\mathbf{V}}_{t,n,3},\\
\dot{\mathbf{V}}_{t,n,6}&\triangleq \frac{\partial\mathbf{V}_{t,n}}{\partial \Delta_d} =  w_t\dot{\mathbf{V}}_{t,n,5},
\end{align}\end{subequations}
where $u_t$, $v_t$, $w_t$ are the normalized offset indices for the $t$-th scatterer defined in Sec. II-D.

Considering the array geometry described in Sec. II-C, the transmit steering vector derivatives take the following forms:
\begin{equation}\label{eq:steer vec derive} \begin{aligned}
\frac{\partial\mathbf{a}_{t,n}}{\partial\theta_0}&=\frac{\partial\mathbf{a}_{\text{x},t,n}}{\partial\theta_0}\otimes\mathbf{a}_{\text{y},t,n} + \mathbf{a}_{\text{x},t,n}\otimes\frac{\partial\mathbf{a}_{\text{y},t,n}}{\partial\theta_0}\\
& = \jmath\pi\chi_n\sin\phi_t
(\sin\theta_t\mathbf{D}_\text{x}\mathbf{a}_{\text{x},t,n}\otimes\mathbf{a}_{\text{y},t,n} \\
&\qquad\qquad\qquad~  - \mathbf{a}_{\text{x},t,n}\otimes\cos\theta_t\mathbf{D}_\text{y}\mathbf{a}_{\text{y},t,n}),\\
\frac{\partial\mathbf{a}_{t,n}}{\partial\phi_0}
&= -\jmath\pi\chi_n\cos\phi_t
(\cos\theta_t\mathbf{D}_\text{x}\mathbf{a}_{\text{x},t,n}\otimes\mathbf{a}_{\text{y},t,n}\\
&\qquad\qquad\qquad\quad~ +\mathbf{a}_{\text{x},t,n}\otimes\sin\theta_t\mathbf{D}_\text{y}\mathbf{a}_{\text{y},t,n}),
\end{aligned}\end{equation}
where $\mathbf{D}_\text{x}\triangleq \text{diag}\{0,1,\dots,N_\text{tx}-1\}$ and $\mathbf{D}_\text{y} = \text{diag}\{0,1,\dots,N_\text{ty}-1\}$. Similar expressions apply to the receive steering vector $\mathbf{b}(\theta,\phi)$. The derivative of the delay-related phase term $f_{t,n}$ with respect to $d_0$ is 
\begin{align}
\frac{\partial f_{t,n}}{\partial d_0}= \frac{-\jmath4 n\pi\Delta_f}{c}f_{t,n}.
\end{align}

The block-diagonal transmit covariance matrix satisfies
\begin{subequations}\begin{align}
 \mathbf{R}_X &\triangleq \frac{1}{L}\mathbb{E}\{\mathbf{X}\mathbf{X}^H\} \\
 & = \frac{1}{L}\mathbb{E}\{\text{blkdiag}\{\mathbf{X}_0\mathbf{X}_0^H,\dots,\mathbf{X}_{N-1}\mathbf{X}_{N-1}^H\}\}\\
 &= \text{blkdiag}\{\mathbf{R}_0, \mathbf{R}_1,\dots,\mathbf{R}_{N-1}\},
\end{align}\end{subequations}
where $\mathbf{R}_n = \mathbf{W}_n\mathbf{W}_n^H$ since $\mathbb{E}\{\mathbf{S}_n\mathbf{S}_n^H\} = L\mathbf{I}_{N_\text{t}+K}$. 
Substituting the above derivatives into the general FIM formula \eqref{eq:Fij} from Sec. III-A, the FIM elements simplify to:
\begin{align}
    \mathbf{F}_\text{gg}(i,j)  &= \frac{2L\sigma_\alpha^2}{\sigma_\text{s}^2}\sum_{t=1}^T\Re\left\{\text{Tr}\left\{\mathbf{R}_X\frac{\partial\mathbf{V}_t^H}{\partial\xi_i}\frac{\partial\mathbf{V}_t}{\partial\xi_j} \right\}\right\}\\
    &= \frac{2L\sigma_\alpha^2}{\sigma_\text{s}^2}\!\sum_{t=1}^T\sum_{n=0}^{N-1}\Re\left\{\text{Tr}\left\{\mathbf{R}_n\frac{\partial\mathbf{V}_{t,n}^H}{\partial\xi_i}\frac{\partial\mathbf{V}_{t,n}}{\partial\xi_j} \right\}\right\}\\
    & = \sum_{n=0}^{N-1}\Re\left\{\text{Tr}\left\{\mathbf{R}_n\mathbf{C}_{i,j,n}\right\}\right\}, \label{eq:Fgg}
\end{align}
where 
\begin{align}
    \mathbf{C}_{i,j,n} = \frac{2L\sigma_\alpha^2}{\sigma_\text{s}^2}\sum_{t=1}^T\dot{\mathbf{V}}_{t,n,i}^H\dot{\mathbf{V}}_{t,n,j}.
\end{align}


\section{FIM Coefficient Matrices under DSM}

Here we present the coefficient matrices required for calculating the FIM under the DSM, as detailed in Sec. IV-B. In the DSM formulation, each scatterer is modeled independently, resulting in derivatives that are non-zero only with respect to the parameters associated with the same scatterer. Specifically, the derivatives of $\mathbf{V}_{t,n}$ with respect to the individual scatterer's parameters $(\theta_t,\phi_t,d_t)$ are expressed as
\begin{subequations}\label{eq:DSM-derivatives}\begin{align}
\dot{\mathbf{V}}_{t,n,\theta_t}&\triangleq \frac{\partial\mathbf{V}_{t,n}}{\partial\theta_t} =  f_{t,n}\frac{\partial\mathbf{b}_{t,n}}{\partial\theta_t}\mathbf{a}_{t,n}^H \!+\! f_{t,n}\mathbf{b}_{t,n}\frac{\partial^H\mathbf{a}_{t,n}}{\partial\theta_t},\\
\dot{\mathbf{V}}_{t,n,\phi_t}&\triangleq\frac{\partial\mathbf{V}_{t,n}}{\partial\phi_t}  = f_{t,n}\frac{\partial\mathbf{b}_{t,n}}{\partial\phi_t}\mathbf{a}_{t,n}^H \!+\! f_{t,n}\mathbf{b}_{t,n}\frac{\partial^H\mathbf{a}_{t,n}}{\partial\phi_t},\\
\dot{\mathbf{V}}_{t,n,d_t}&\triangleq\frac{\partial\mathbf{V}_{t,n}}{\partial d_t}  =  \frac{\partial f_{t,n}}{\partial d_t}\mathbf{b}_{t,n}\mathbf{a}_{t,n}^H.
\end{align}\end{subequations}
The steering vector derivatives in \eqref{eq:DSM-derivatives} share the same structure as those derived under the PSM in Appendix A, with the parameter evaluation performed individually at each scatterer's coordinates $(\theta_t,\phi_t,d_t)$. Similarly, the delay derivative is computed as $\partial f_{t,n}/\partial d_t = -\jmath 4\pi n \Delta_f f_{t,n}/c$. Due to the independent scattering assumption inherent in DSM, the cross-terms in the FIM involving different scatterers vanish upon averaging over the independent scatterer reflectivities. Consequently, the coefficient matrices used in the FIM calculation, as introduced in Sec. IV-B, exhibit a block-diagonal structure:
\begin{align} \mathbf{C}_{i,j,n}^\text{DSM} \!=\!  \begin{cases}
\frac{2L\sigma_\alpha^2}{\sigma_\text{s}^2} \dot{\mathbf{V}}_{t,n,i}^H\dot{\mathbf{V}}_{t,n,j}, & \text{if } i, j \text{ correspond to scatterer } t, \\
\mathbf{0}, & \text{otherwise},
\end{cases}
\end{align}
where the indices $i,j$ correspond to the scatterer-specific parameters defined for $\bm{\xi}_\text{DSM}$ in \eqref{eq:DSM paras} of Sec. IV-B. 


\end{appendices}

\end{document}